A FOOLISH CONSISTENCY? ALIGNING INTERFACE OBJECTS HINDERS LOCATION RECALL AND MAY INDUCE COLLINEAR ERRORS

Peter Zelchenko, Li Xiangqian, Fu Xiaohan, Alex Ivanov, Zhenyu Gu

During our nearly constant use of digital devices, perhaps our most frequent need is to visually identify icons representing our content and invoke the actions to manipulate them. Almost since the inception of user interface design in the 1970s, with rare exception it has become the tendency for programmers to prescribe the arrangement these things in uniform rectilinear rows and columns. This was imported from theories for print design and ultimately brought into widespread practice for graphical user interfaces (GUIs). Whether consistent rectilinearity actually does better than less rectilinear arrangements to maximize selection efficiency has not been challenged on considerations of speed or any other measure. In a series of four experiments, we explore how alignment may in fact discourage easy recallability of screen object locations and hence increase search intensity. A second objective is to present a methodological model where we deliberately attempt to begin with psychophysical cognitive evidence at the environmental schematic low end (beginning with contextual cueing paradigm), then move progressively upwards in naturalism in the experiments to something that approximates actual human work at the higher end, all along attempting to keep one important environmental property constant. Two experiments using contextual cueing paradigm confirm that collinearly aligned arrays do not encourage recallability of location, while noncollinear arrays appear to create traces that can be recalled automatically. Two other experiments give further demonstration of explicit recollection and location recall, showing that collinear arrangements may in fact induce location recall errors to neighboring collinear objects. We discuss surrounding theoretical, historical, and practical questions.

## 1 Introduction

During our nearly constant use of digital devices, perhaps our most frequent need is to visually identify icons representing our content and invoke the actions to manipulate them. Conveniently locating controls and content is one of the most widely studied concerns in human-computer interaction (HCI) design. To address this, almost from the inception of user interface design in the 1970's, with rare exception it became the tendency for programmers to prescribe the arrangement these things in uniform rectilinear rows and columns. This was brought into standard practice for graphical user interfaces (GUIs) not long after.

Whether consistent rectilinearity actually does better than less rectilinear arrangements to maximize selection efficiency has not, to our knowledge, been challenged on considerations of speed or any other measure, let alone whether such studious uniformity might cause confoundment in object location recall among other nearby objects. Anecdotally, this seems plausible. There have been efforts in cognitive and environmental psychology on the one hand, and HCI on the other, to examine questions about search efficiency, but theoretical and applied sciences have not been closely linked. Attention and selection, perceptual and semantic object memories, eye tracking, cognitive load, affordance theory, and other areas all have been built up through many years of scientific evidence, but when taken up by developers and designers as they often are, they seem superficial and often merely speculative. The bridges are probably not very long ones; however, it has been unclear how to stretch them across the interdisciplinary gap. Perhaps the basic notion that frequent visual searching through lists should be the default on modern devices should be challenged on one of the most unanimous certainties from cognitive science: that serial feature search is cognitively costly. The many brief microsearches up and down rectilinear arrays is such a familiar activity that it seems interaction designers have ignored this point. It serves as the design plan for acquiring nearly every object today in our organized world (Figure of Apple guidelines). Apple's guidelines state, "When content isn't consistently spaced, it no longer looks like a grid and it's harder for people to scan." This general proposition serves to guide the default screen object properties for all of today's standard graphical interfaces. And yet Gestalt theory and cognitive psychology both treat collinearity (Gestalt "good" continuation) as the essential building block for primitive shapes to give rise to emergent structures. These configural structures challenge attention in unexpected ways. Put in another way, this means that collinear objects radiate imaginary lines that urge the visual system to follow them – and these imaginary lines are often stronger even than contrasting colors. And, unlike contrasting colors, the imaginary lines are often unintended. In particular, they *may* benefit browsing, but they may also interfere with visual search, creating distractive forces that cause the eye to stray widely from a target, interfering with normal object selection processes. Hence our devices, as well as our urban environments, do not benefit from adequate study of cognitive psychology: instead, they remain tied to tried-and-true rectilinear design patterns that may be at cross-purposes to human cognition. Humans are thus often made to adapt to developers' ingrained impulses, and psychologists are helpless to make useful contributions.

Our series of experiments has two primary objectives. The chief one is to explore how alignment may in fact discourage easy recallability of screen object locations and hence increase search intensity. Surprisingly, after 40 years of the GUI and society's reliance on interface interactions, this is is something that we have not seen explored. No study that we know of has directly addressed rectilinearity as a potential problem for cognition and memory.

A second objective is to present a methodological model where we deliberately attempt to begin with psychophysical cognitive evidence at the environmental schematic low end, moving progressively upwards in naturalism in the experiments to something that approximates actual human work at the higher end, all along attempting to keep one important environmental property constant. This should help establish a chain of custody of environmental affordances and evidence from the perceptually primitive level to practical, everyday interaction – an interdisciplinary model that we feel would benefit human interface design.

# 2 Literature review

## 2.1 Gestalt and part-whole hierarchy

Theories for perceptual grouping are seeing keen interest today. Recent attempts to model perceptual grouping strategies have shown some success in using Voronoi (Machilsen, Wagemans, & Demeyer, 2016) and Bayesian (Jäkel, Singh, Wichmann, & Herzog, 2016; Wagemans et al., 2012) spacing algorithms, as just two examples. Such work serves both to offer mathematical models for artificial generation of space as well as to attempt to give more accurate accounts of human perception in the interaction age. However, the principle that spatial proximity serves preattentive interpretation of perceptual structures has already long been acknowledged in cognitive psychology (Kimchi, 2009, 2015; Navon, 1977; Pomerantz, Sager, & Stoever, 1977). Perceptual organization is largely preattentive and bottom-up (Kimchi, 2009; Treisman, 1982). The whole-part principle of the Gestaltists took form in the cognitive theories as *holistic* or *configural* properties or features (Wagemans et al., 2012, p. 1223). In the case of global cues, the whole-part problem touches not only upon individual features which constitute closure objects, but also applies to local grouped clusters and arrangements of objects, such as icons on a screen. Thus, for both objects and hierarchical groups of objects, interstitial space among subparts is relevant both to grouping as well as object discrimination.

## 2.2 Attention, selection, and effort

Processes that require focused attention, including short visual feature searches, are serial, slow, and require controlled attention; whereas processes that do not require focused attention are generally parallel, fast, and automatic (Hasher & Zacks, 1979; Schneider & Shiffrin, 1977; Shiffrin & Schneider, 1977). Selection that does not require attention is sometimes said to be effortless (Logan, 1992; Rensink & Enns, 1995); however, the meaning, extent, and locale of so-called effort is not always clear (see, e.g., Wolfe & Horowitz, 2017, p. 6). The claim has been attenuated to mean that in such cases both attendance to objects and concomitant effort are unconscious, of substantially less effort than those involving focused attention, generally not calling for stimulus-reward reinforcement (Anderson, 2015; Theeuwes, 2018). That feature search requires attention and effort, and that subconscious parallel filtering requires little or perhaps no conscious focused attention, lower cognitive load, and is thus effectively automatic, are, however, generally accepted. Likewise, the concepts of top-down and bottom-up processing have been challenged (Awh, Belopolsky, & Theeuwes, 2012). Some researchers have begun to clarify the difference between top-down or goal-directed attention and bottom-up attention with what they now call history-based or habitual attention, so as to broaden the concept of mere bottom-up claims of attention to that of attention guided by a combination of goals, object saliency and memory of previous selections (Awh et al., 2012; Y. V. Jiang, 2018; Theeuwes, 2018). All this points to the greater role and value of memory in interaction. Since visual search is cognitively costly, information should be designed to reduce the need for controlled feature search to the extent possible for selecting objects, instead encouraging the most automatic means available. With history-based attention, prior long-term memories influence automatic selection.

### 2.2.1 Instance theories and the contextual cueing paradigm

Tying attention more closely to memory has helped to explain how memory contributes to automatic cognitive processes, and it has also reinforced claims establishing feedback relationships between memory and perceptual stimuli (Konkle, Brady, Alvarez, & Oliva, 2010; Tseng & Jingling, 2015). Implicit or incidental learning proposals beginning in the 1970's have contributed to a large number of unified theoretical models (Goujon, Didierjean, & Thorpe, 2015). Many proposals have surfaced over several decades, and neuroscience also has been helping to sort out their discrepancies. As part of this trend, instance theories of attention hold that every exposure to a stimulus is stored and leads to memory reinforcement. Instance theories are part of an interconnected family of related theories and models, including those concerned with statistical and implicit learning, associative learning, exemplar learning, episodic memory, global matching, and other lines, all attempting to explain how stimulus perception is organized into established memories (e.g., Hintzman, 1984; viz. Jamieson, Crump, & Hannah, 2012; Palmeri, 1997; Raaijmakers & Shiffrin, 1992; Shriffrin & Steyvers, 1997). It also claims theoretical links to semantic memory, judgment, and problem-solving (Logan, 1988). Instance theories are recently playing a central role in a general model explaining episodic memory and associative learning (Jamieson, Johns, Vokey, & Jones, 2022).

Logan's instance theory (1988) places particularly strong reliance on memory. It describes automaticity entirely as a phenomenon of memory, setting aside resource-dependent approaches to automaticity and instead equating automaticity with the fact of memory retrieval: "performance is automatic when it is based on direct-access retrieval of past solutions from memory" (Logan, 1988, p. 493). It provides that encoding is obligatory and unavoidable, merely requiring attention, and that retrieval is likewise invoked by mere attention. However, it emphasizes that the quality and quantity of attention are relevant to the quality of the encodings (Logan, 1988, p. 494).

Studies since the 1970s began demonstrating increased performance when targets appeared in areas where they had appeared most often before. This location probability, also called probability cueing, was refined in the late 1990s with the discovery that spatial arrays could also trigger probability cueing for targets within the array. This led to the further discovery that target locations may be learned implicitly due to the spatial properties of neighboring array objects. A unique method was developed to demonstrate this effect, called contextual cueing (Chun & Jiang, 1998; for a detailed summary, see Jiang, 2018). Contextual cueing (CC) assumes the obligatory encoding and retrieval aspect of Logan's instance theory. Demonstrating implicit learning, the basic CC experiments (Chun & Jiang, 1998) use classic psychophysical serial-search tasks, whose antecedents were originally used in seminal attention and selection theory development (Duncan & Humphreys, 1989; Wolfe, Cave, & Franzel, 1989). These were designed as T-among-L target/distractor arrays presented in rapid succession; in CC-paradigm experiments, the manipulation is a small set of so-called "old" trial displays randomly shuffled and re-inserted within a much larger set of never-repeating "new" trial displays (Figure 1). Superior reaction times (RTs) for "old" displays after even brief training demonstrate both the implicit learning from local contextual information and the richness of the contextual capture. CC demonstrates that an associative learning relationship exists between targets and their distractors, and that an implicitly encoded memory for the target locations is reinforced by this relationship (Goujon et al., 2015, p. 526). In CC, qualities of local distractor context contribute to location recall of the target (Annac, Zang, Müller, & Geyer, 2019). In other words, it makes evident that the spatial relationships among targets and nearby distractors, and not solely object-internal features, are relevant to object location recall. CC was shown early on to survive interference from secondary tasks, suggesting that its working memory demands are low (Vickery, Sussman, & Jiang, 2010). This finding has subsequently been extensively revisited, with an apparent consensus that CC does not rely on visual or spatial WM resources during the learning phase, but it may during the testing phase (Amsalem, Sahar, & Makovski, 2023; Annac et al., 2019; Goujon et al., 2015; Manginelli, Langer, Klose, & Pollmann, 2013; Travis, Mattingley, & Dux, 2013; Vicente-Conesa, Giménez-Fernández, Shanks, & Vadillo, 2022; Yang, Chen, He, & Merrill, 2020). Recently, the findings were extended to apply to elementary school students (Yang et al., 2020). In addition, Annac and colleagues (2019) found that performing CC tasks under WM load actually led to an *increased* effect of CC. Whereas during pattern learning, contextual cueing depends on the deployment of selective attention to relevant areas of the scene (Jiang & Chun, 2001; Jiang & Leung, 2005; see Conci, 2009).

Both theoretical and applied research in CC to date has examined certain collinear interrelationships among target and distractor locations but has not studied the difference between collinearity and other arrangements among targets and distractors. To our knowledge, only one CC study has been done that studied device displays, but it focused on automatic remapping of rectilinear displays when switching between portrait and landscape modes. Researchers found substantial differences in recall depending on how smartphone displays are automatically rearranged (Shi, Zang, Jia, Geyer, & Müller, 2013).

Table 1. Canonical Form of a Contextual Cueing Experiment

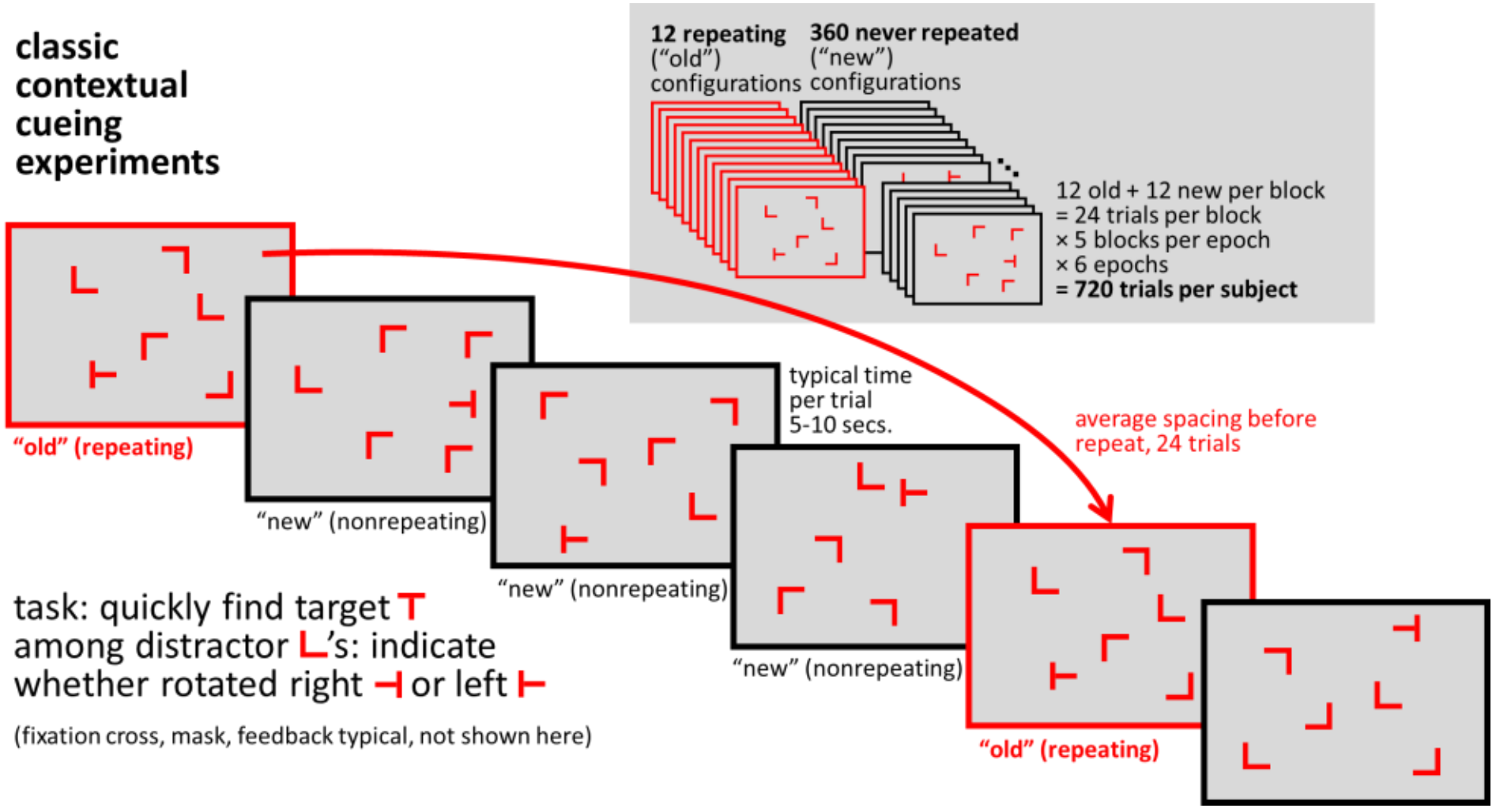

Figure 1. Contextual cueing paradigm shows contribution of memory, as reaction times improve for "old" (repeating) contexts but not for "new" (never repeated) contexts.

## 2.3 Recognition memory

Recognition memory – the ability of the normally functioning brain to recall that a stimulus was encountered in the past – is inherently associative in nature. When individuals recall a stimulus, it is likely tied to certain other attributes, including spatial location (Schulman, 1973; Zechmeister & McKillip, 1972). Recognition memory is believed to be automatic (Hasher & Zacks, 1979), although the this has been hotly debated. automatic processes should decrease cognitive demand more effectively than effortful processes. Of relevance to our work is that CC subjects do not report any greater familiarity with "old" displays than the "new" displays or than chance, immediately after the experiments. Somewhere along the "chain of custody" in our studies between CC-based encoding in LTM and slow rehearsal there must occur a transition in recognition memory of worked displays from nonfamiliarity to familiarity.

## 2.4 GUI studies

Efforts in HCI to examine graphical user interfaces today hinge on a triumvirate of practice: the research questions relate to saliency, as applied to web sites, and predominantly analyzed with eye tracking (Hicks, Cain, & Still, 2017; Still, Hicks, & Cain, 2020). With CCD arrays dropping in price and increasing in resolution, beginning in the early 2000's eye tracking became a favored, reliable tool for determining attentional focus, and the canonical "F" pattern[1] for websites was identified and confirmed through various studies (Djamasbi, Siegel, & Tullis, 2011; Djamasbi, Siegel, Tullis, & Dai, 2009; Shrestha & Lenz, 2007; see Shrestha, Owens, & Chaparro, 2009 for a review). Faraday (2000) popularized a simple set of rules proposing various hierarchical categories of web "entry points" (e.g., large key objects, priority positions, etc.) for saliency priority. Although supported by only limited evidence and later shown to rest on weak assumptions (Still, 2018; Still & Masciocchi, 2012), these rules are still widely used today as a standard reference for practice. In the HCI attention literature touching on memory, there has been some support for claims that attending to salient objects in attractive, spatially open layouts fosters greater website recall and preference (e.g., Sutcliffe & Namoune, 2008). More recently, work has been done to evaluate the effectiveness of visual enhancements of online advertising, exploring object saliency due to such devices as strategic use of color (Hamborg, Bruns, Ollermann, & Kaspar, 2012; Rumpf, Boronczyk, & Breuer, 2019), animation and color (Breuer & Rumpf, 2015), and possibly personalization (Stiglbauer & Kovacs, 2018). Where location and recall are discussed in these studies, they tend to be limited to confirming or deliberating what is known from eye-tracking studies regarding attentional priorities of humans viewing web content. Spatial openness is relevant to the present study, as the more ample negative space could serve location recall. However, recall of salient objects does not equal recall of their location. Since much of the literature is geared toward marketing tasks, recall in HCI has been focused on websites, aesthetic qualities, and target semantic recall (but not location recall). It does not otherwise consider object location per se but rather whether a particular brand or its salable product is more memorable than others due chiefly to some accord between an object's saliency and its other available episodic factors, including emotions.

[1] This "F"-shaped eye-tracking heatmap phenomenon was discovered some years earlier by marketers. It was labeled "[Google's] golden triangle," due to the consistency of the rotated triangular-shaped heatmap in the upper-left corner of the screen. The authors indicated that sites shown in this small region in search-engine results were more likely to be in the area of the heatmap. See Hotchkiss, G., Alston, S., & Edwards, G. (2005). "Eye tracking study: Released by Enquiro, I-Tools, and Did-It."

It is worth noting here that in general in the HCI literature, object saliency and the so-called "pop-out effect" tend to be interpreted to mean any design device that might especially attract the eye, such as additional animation or more vibrant color. This is not to claim that the usage is wholly inaccurate, only that the term is less formally applied in HCI as design advice and that it remains to be demonstrated whether all experimental results regarding saliency and pop-out among HCI theorists and practitioners are in fact all describing the automatic, parallel-processed attentional conjunction phenomenon originally identified in feature integration theories of attention (Treisman & Gelade, 1980).

Researchers have shown that interesting regions within a web page are more likely to be remembered by a user (Sutcliffe & Namoune, 2008, p. 18). However, immedately there arises a conceptual conflict: These same researchers argue the upsides of columnar structure as beneficial for "framing attention" and for promoting "scanning." They further state that uncluttered layouts "reduce competing stimuli for user attention" (Sutcliffe & Namoune, 2008, p. 18). This may be correct for some past and even future designs. However, as we hope to demonstrate, it may be exactly these uniform structures and the components of these structures whose similarity would neutralize any spatial uniqueness, and so they might not encode unique traces for location-memory retrieval.

## 2.5 Good continuation, collinearity, and rectilinearity

Collinearity – the modern incarnation of Gestalt "good" continuation – is a chief contributor to emergent features that cause a collection of objects to coalesce into a perceptual whole (Pomerantz et al., 1977; Wagemans et al., 2012). Can the essential local perceptual organizing feature of collinearity also be the very same thing that drives global confusion? Most animal and machine visual systems bias for collinearity as a basic function for discriminating objects from backgrounds. Humans have a further built-in bias for rectilinearity (Kubovy & Wagemans, 1995)(Brooks, 2015; Tversky, 1981, 1990; Wagemans et al., 2012, p. 1221) and even for correcting diagonal contours to 45 degrees (Schiano & Tversky, 1992; Tversky, 1981, 1990). Jingling and Tseng (2013; 2015) used columns composed of short collinear bars as distractors in a target-selection activity, and found the collinearity to impair visual search for local elements. Contextual cueing experiments have also begun to show some allied evidence of this phenomenon. In work similar to that of Jingling and Tseng, Conci and colleagues (2013; 2009) used the "L" distractors typically used in CC studies to construct partially formed square regions – reminiscent of Gestalt illusions (Figure) – to demonstrate how segmentation by these regions serves simultaneously to structure attention and in the process to hamper learning. Kimchi and colleagues (2016) showed that the greater the saliency of such emergent collinear structures, the higher the RT, and consequently the more stimulus-driven attentional capture by the distractor. It would stand to reason that the collinear matter so prevalent in today's user interfaces could create emergent patterns and contours that could strongly influence attention and lead to impaired search as well as memory confoundment. It also follows that some kind of noncollinear arrangements could reduce this effect and possibly provide further benefits.

## 2.6 Our experiments

The general class of topographic spatial recall experiment designs that we relied on is a configural hybrid of CC and higher-level experiments. For the higher-level designs, we studied early location-recall experiments by Mandler et al. (1977) and a derivative by Pezdek et al. (Pezdek, 1983; Pezdek, Roman, & Sobolik, 1986a). These sparse displays rather predate the era of the cluttered graphical interface, but they do evoke the minimalist design intent of displays as envisioned by Xerox PARC and the Macintosh, which has seen a resurgence in recent years. We also examined work in this line using similar experiments from Kessels et al. (1999) and Postma et al. (2004; 2008) on recall and forgetting, and Hicks et al. (2017) on the influence of clutter.

Based on various findings and arguments in the related literature, we pursued the following research questions:

a. How does the presentation of collinear screen matter affect object location recall?
b. Can the contextual cueing paradigm detect a difference in object-location learning between collinear and noncollinear screen matter?
c. Can effects of such cognitive phenomena be traced in a responsible way up the ecological path, using an evidentiary chain of custody leading from psychophysical evidence to more naturalistic evidence?

# 3 Experiment 1

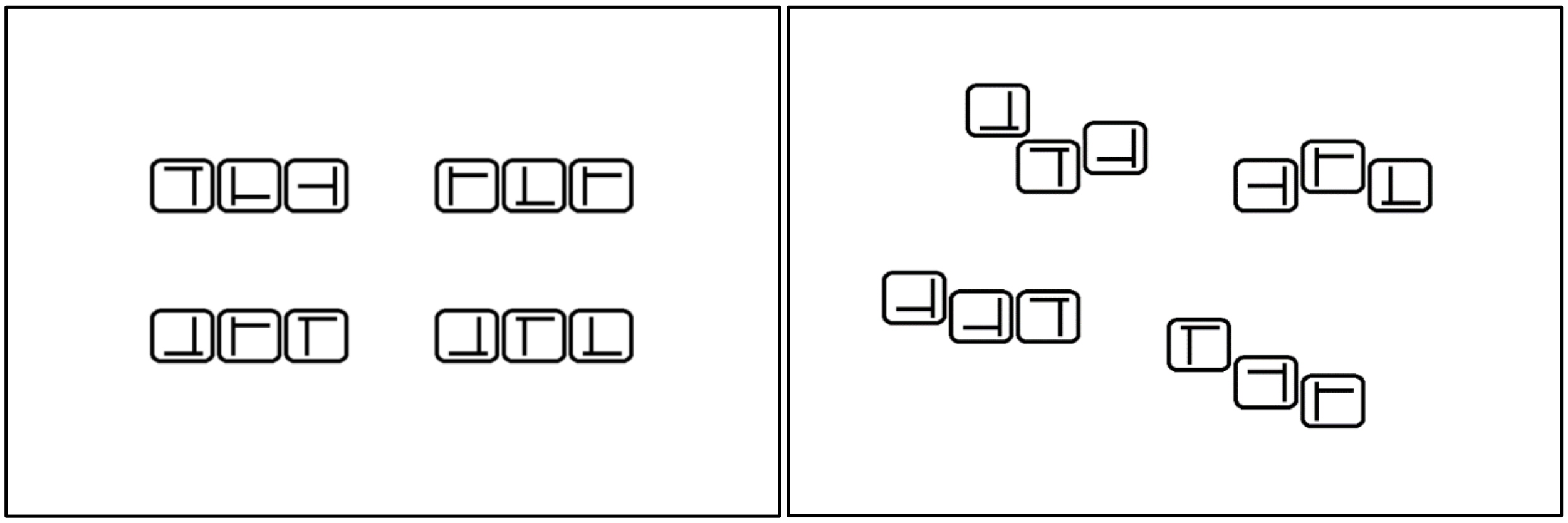

Figure 2. Representative displays used in the collinear (left) and noncollinear conditions of Experiment 1.

Our intention was to begin at a low perceptual level and move up progressively to greater naturalism. We used the CC paradigm in Experiment 1 to compare conventional collinear arrangements of objects versus deliberately noncollinear icons. Our argument was that differenes in CC effect would help demonstrate that participants are able to implicitly remember object locations in noncollinear layouts better than collinear ones. When locating target items, they would no longer rely much on visual feature search but would more easily implicitly recall the target's spatial location, which after a short time should facilitate efficient object access.

We made the following predictions:

1. Due to contextual cueing, after training, in the final three epochs of training, response latency for both conditions should be greater in "new" screen trials, while that of "old" screen trials should be relatively small ("CC effect").
2. Participants in the noncollinear group should experience a larger CC effect than those in the collinear group.
3. Participants in the noncollinear group should show lower error rates (ER) than collinear group participants.

## 3.1 Method

### 3.1.1 Research ethics disclosure

The approval of ethical standards for Experiments 1, 2, and 3 was given by the Ethics Committee of Fudan University. All participants provided informed consent by checking a box prior to the experiment.

### 3.1.2 Design and Participants

Studies using contextual cueing have typically recruited only 4 to 15 participants (e.g., Chun & Jiang, 1998, 1999; Chun & Phelps, 1999; Zinchenko et al., 2018). Our stimuli design deviated from many CC experiments in that we added a standard border (see 3.1.3 below). Given the concern that this might inhibit encoding of local features, in the statistical power analysis (Faul, Erdfelder, Lang, & Buchner, 2007), we expected to have a small effect size (i.e., $f = 0.15$). In addition, our use of CC was not to measure the CC effect alone in perceptual measurements, but to compare two different approaches to spatial arrangement. We would investigate the contextual cueing effects in a three-way ANOVA with repeated measures and within-between interactions. The results of the analysis suggested an optimal sample size of $N = 62$ for $\alpha = .05$ and power = .80, with 31 participants in each layout group.

A total of 70 participants from Fudan University took part in Experiment 1. Four students from the collinear group and four students from the noncollinear group did not achieve a sufficiently low error rate in each epoch (> 25%) and had to be replaced. Ultimately, we established a balanced sample of 31 participants in each layout group. Each participant received 50 RMB (≈ USD$8) for taking part.

### 3.1.3 Apparatus, Stimuli, and Task

Experiment 1 was programmed using the PsyToolkit online platform (Stoet, 2020, 2017). Our goal was to explore concerns among users of digital devices with objects equilinearly spaced on their centers. We sought to design a template for experiments at all levels that would approximately emulate existing design practice. Accordingly, our master template for all experiments would vary collinearity in the same way, while advancing up the ladder of realism by exploring differences in task and in number of objects.

#### *3.1.3.1 Stimuli generation*

Internal features of stimuli for Experiment 1 were chosen as “T” target among “L” distractors, dating to Beck (1966; Duncan & Humphreys, 1989) and traditional in CC experiments (e.g., Bergmann, Koch, & Schubö, 2019; Chun & Jiang, 1998; Y. V. Jiang & Chun, 2001; Travis et al., 2013). Although the basic concept of Experiment 1 was similar to these studies, practical aspects influenced our decisions. First, to more closely resemble computer or smartphone interfaces, we added a border to each letter (see Figure 2). Secondly, we had originally planned to place stimuli in collinear arrays with fully monolinear grids all with equal *x* and *y* variation, in the noncollinear condition staggering objects randomly from those centers. However, arguments favoring collinearity as superior for search speed might admit that a fully monolinear array could give it an unfair advantage. Arguments favoring noncollinearity might admit that even a small difference in space between two objects could induce some perceptual grouping unavailable to the collinear condition. Accordingly, we maintained a consistent level of hierarchical grouping by threes. This, it was felt, would test linearity but also offer comparable hierarchical chunks for both conditions, under the assumption that such grouping is automatic (Moore & Egeth, 1997) and should by itself offer no advantage to either condition. Therefore, for all experiments, we chose to cluster elements for both collinear and noncollinear spatial conditions in chunks of three objects.

Linearity for the collinear condition was defined as the triplet maintaining alignment in the *x* line, and sets of triplets being equidistantly spaced in *y*. From this perfect linearity, noncollinear arrays were generated automatically and randomly from the set of all possible triplet arrangements given one object occupying the center and two others in edge contact and in orbit around that central object. From this, a configural quasi "entropy" for triplets was defined as the extent to which a triplet's bounding rectangle differed from the neutral aspect ratio of a horizontally collinear triplet. This meant that triplets forming diagonal or vertical lines would be assigned the highest entropy and triplets forming near-triangles were assigned a middle level of entropy. The computation was made by taking the area of the triangle formed by each triplet based on the Heron semiperimeter of the three objects, as well as by taking into account the angle of slope created by the longest side of the triangle. The intent was to maintain relatively low configural entropy deviating from collinear, with triplets still maintaining orbital contact; accordingly, odds bins were computed from this assumption that favored low levels of entropy, and triplets were selected from the lower end of these odds. Spacing between triplet groups was determined by a random offset that maintained noncontact among contiguous triplets. Finally, the empty arrays were populated with interior features. Four fixed sets of stimuli (360 "new" + 12 "old" + 12 practice + 12 testing) were used for all participants, although presented in a different random order for each participant (see below).

#### *3.1.3.2 Task*

For both CC experiments, the task was essentially that given by Chun & Jiang (1998), the two differences being the design of the trial displays and stimuli and the fact of the collinear versus noncollinear conditions given to the two separate groups. The goal was both to measure any CC effect as well as compare the effects between the two design conditions. Participants were given a total of 720 trial displays, divided evenly among 30 blocks of 24 trials each. (In data analysis, each block would be grouped into six epochs of 5 task sets each, or 120 trial displays for each epoch.) Each block of 24 trial displays was randomly shuffled and participants were presented 12 so-called "new" trial displays (never before and never again seen), mixed with 12 so-called "old" trial displays (inserted among the 12 "new" displays). Therefore, in the classic CC experiment, a total 360 "new" displays are mixed and presented with an equal number of 12 "old" displays presented 30 times each, thus also amounting to 360 exposures, in a random but systematic mix. CC arises when reaction time (RT) decreases with "old" arrangements only, and yet viewers confirm no memory of any arrangements or target locations. In both conditions, the target occupied a different position in the array for each of the 12 "old" displays and also appeared in each position an equal number of times in the set of "new" displays. During the experiment, a training task was first given, using 12 never-again-used configurations. In the main task, in each trial, first a fixation cross was flashed three times in a random position near center screen over 450 ms. Next, a trial display appeared and remained for up to 5 s, during which period the participant would indicate having found the target by pressing the "q" key on the keyboard if the target was pointing left, "p" if pointing right. If 5 seconds expired with no response, the next trial would begin. Corresponding correct or incorrect feedback were flashed for a total of 250 ms. (A trial scheme including the time sequence is provided in Figure 3.) A rest of 30 seconds was offered after every block of 24 trial display items. After all 720 exposures over the 30 blocks, surprise control display recognition and object location recall tests of the 12 "old" mixed with 12 random, brand-new "new" displays was presented as only 24 blind arrays in random sequence. Subjects were first shown each display arrangement and asked whether they had any recollection of it. Next, they were asked to indicate in which square the target had appeared for each of these same displays. However, this was done with the noncollinear condition only, as the blind arrays for the collinear condition would all be identical.

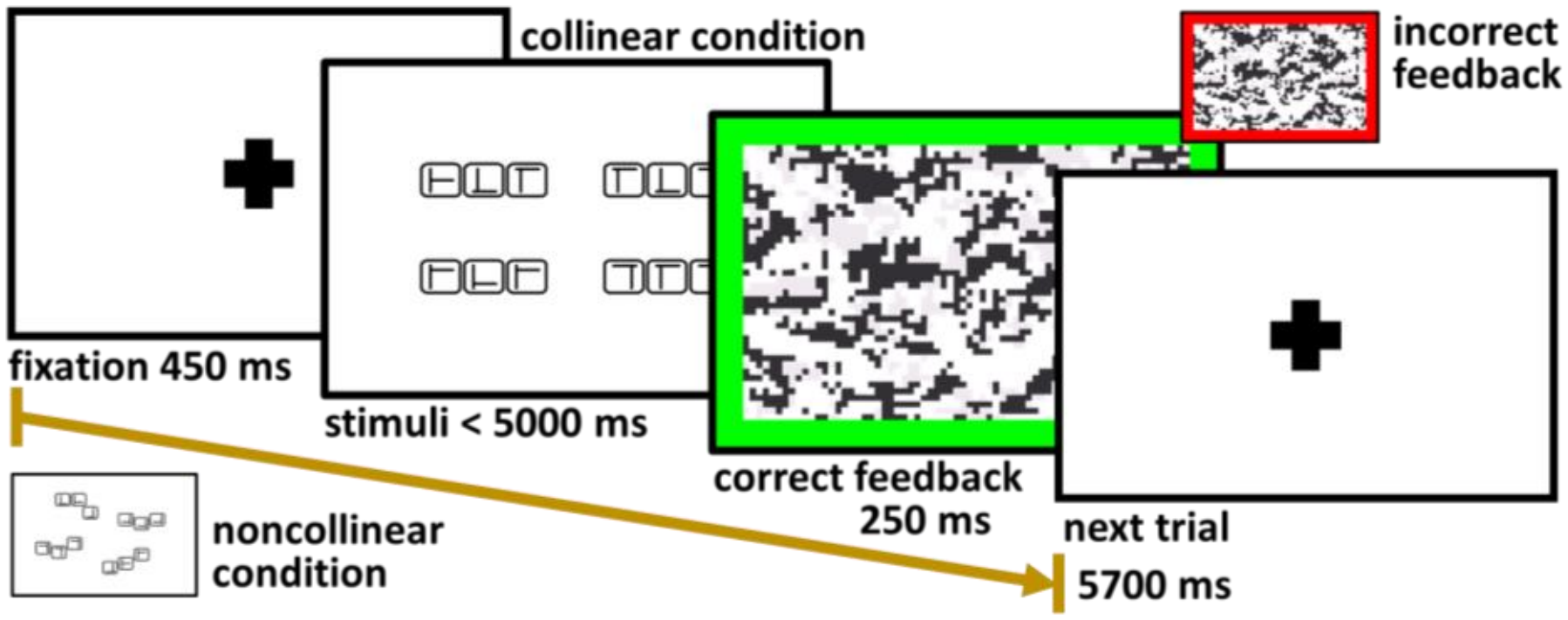

Figure 3. Trial scheme for Experiment 1. Experiment 2 was identical except for the choice of stimuli, which were artificially designed butterflies.

### 3.1.4 Procedure

Students first signed up with a research assistant who randomly assigned each participant to the collinear or noncollinear layout group. Participants were alone in a room with subdued lighting selected to limit distractions, seated at a distance of 50 cm from the display. The working area of the display subtended 18.6° x 12.5° (18 cm × 12 cm) and objects subtended 2.5° of visual angle. Participants clicked a button to enter the experiment. Before starting, they were asked to click a box to confirm that they had agreed to participate in the experiment and informed that they had the right to withdraw at any point. After this step, participants read on-screen instructions and then completed the assigned visual searching task (collinear or noncollinear) and the recall task. Participants who completed the visual searching task were sent compensation via Alipay.

### 3.1.5 Data Analyses

In the present study, data were analyzed with the statistical software R Version 4.2.0 (2022). In the following analyses, all training trials were excluded. Similar to previous studies on CC effects (Chun & Jiang, 1998), we aggregated five blocks into one epoch (e.g., blocks 1-5 = Epoch 1). Participants had to complete 30 visual searching blocks, hence we aggregated them into six epochs. Incorrect trials were excluded from RT analyses.

## 3.2 **Results**

Two identical three-way ANOVAs were conducted to compare the mean RTs and ERs between and within conditions. The two within-subjects factors were epoch (Epoch 1 to Epoch 6), and context ("new" or "old" display). The between-subjects factor was arrangement (collinear or noncollinear). Mean RTs, ERs, and corresponding *SD*s for each trial condition and participant group are shown in Figure X and listed in Appendix B, respectively. Results of the ANOVAs are summarized in Table X and illustrated in Figure 4.

Chun and Jiang, in their original paper (1998), left a useful measurement standard for CC. Our approach for computing the CC effect likewise involved aggregating reaction time data from the last three of six epochs. A so-called "contextual cueing effect" is derived by comparing the differences in mean reaction time (RT) between "old" screen trials and "new" screen trials across those last three epochs. Although the fascinating point of CC paradigm is that it tends to begin showing clear indications of this CC effect even within the first and second epochs (i.e., often fewer than 10 exposures), data from the first three epochs are excluded from the CC effect calculation.

For the RT analyses, all three main effects were significant. Participants' RTs for both conditions decreased gradually from Epoch 1 to Epoch 6. Overall, participants in the collinear layout group (*M* = 1914 ms, *SD* = 292) had shorter RTs than participants in the noncollinear layout group (*M* = 2046 ms, *SD* = 290). In aggregate, participants had shorter RTs in “old” screen trials (*M* = 1945 ms, SD = 299) than in “new” screen trials (*M* = 2016 ms, *SD* = 294). Three significant main effects were moderated by three two-way interactions, and the three-way interactions were also significant. In order to better interpret the results, we first further tested the interaction between context and epochs in each layout group separately. The results showed that the interaction between context and epochs was significant only in the noncollinear group, $F(5, 150) = 7.26$, $p < .001$, $\eta^2_p = .194$, and not significant in the collinear group, $F(5, 150) = 1.85$, $p = .105$.

We further aggregated participants’ RT in the last three epochs. Post-hoc pairwise comparisons were always adjusted for multiple comparisons after Holm (1979). A contextual cueing effect was shown to be significant in both the noncollinear condition (new vs. old = 2007 ms vs. 1835 ms, $p < .001$, $d = +0.76$), and the collinear condition (new vs. old = 1804 ms vs. 1761 ms, $p = .003$, $d = +0.23$). However, participants in the noncollinear condition had larger context cueing effect (new – old = 171 ms) than participants in the collinear condition (new – old = +43 ms), $p < .001$, $d = +1.43$.

In order to compare performance between the two experimental groups, further pairwise comparisons were applied. In each epoch, the average RT of the collinear group “old” trials served as a baseline. We then compared the RT of “old” and “new” screen trials from the noncollinear group with that baseline (Figure 4(b)). For the noncollinear group “new” trials, the results showed that a significant RT delay emerged from Epoch 3, and the difference remained significant through Epochs 4, 5, and 6 (all $p < .005$). For the noncollinear group “old” trials, the results showed a significant RT delay emerging at Epochs 3 and 4. However, the delay was reduced in Epochs 5 and 6 and became insignificant again (all $p > .05$).

For the error-rate (ER) analyses, all three main effects were significant. Participants' ER decreased gradually from Epoch 1 to Epoch 6. Overall, participants in the collinear layout group ($M$ = 3.68%, $SD$ = 3.64) had shorter RT than participants in the noncollinear layout group ($M$ = 5.71%, $SD$ = 4.59). Moreover, participants tended to have lower ERs in old screen trials ($M$ = 4.19%, $SD$ = 4.22) than in new screen trials ($M$ = 5.21%, $SD$ = 4.25). The two-way interaction between layout group and epoch was significant. Overall, the ER difference between the noncollinear group and the collinear group decreased gradually from Epoch 1 to Epoch 6. Pairwise comparisons suggested that the decrease in ER difference between the noncollinear new trials and the collinear group (including both old and new trials) remained significant in all six Epochs (all $p < .05$). However, the ER difference between the noncollinear "old" trials and those of the collinear group was only significant in Epoch 1 and Epoch 2 (both $p < .05$), but not significant in the remaining four epochs (all $p > .05$). The ER difference between the collinear group and noncollinear group was significant in the Epochs 1, 3, and 4 (all $p < .05$), but the difference was insignificant in Epochs 2, 5, and 6 (all $p > .05$).

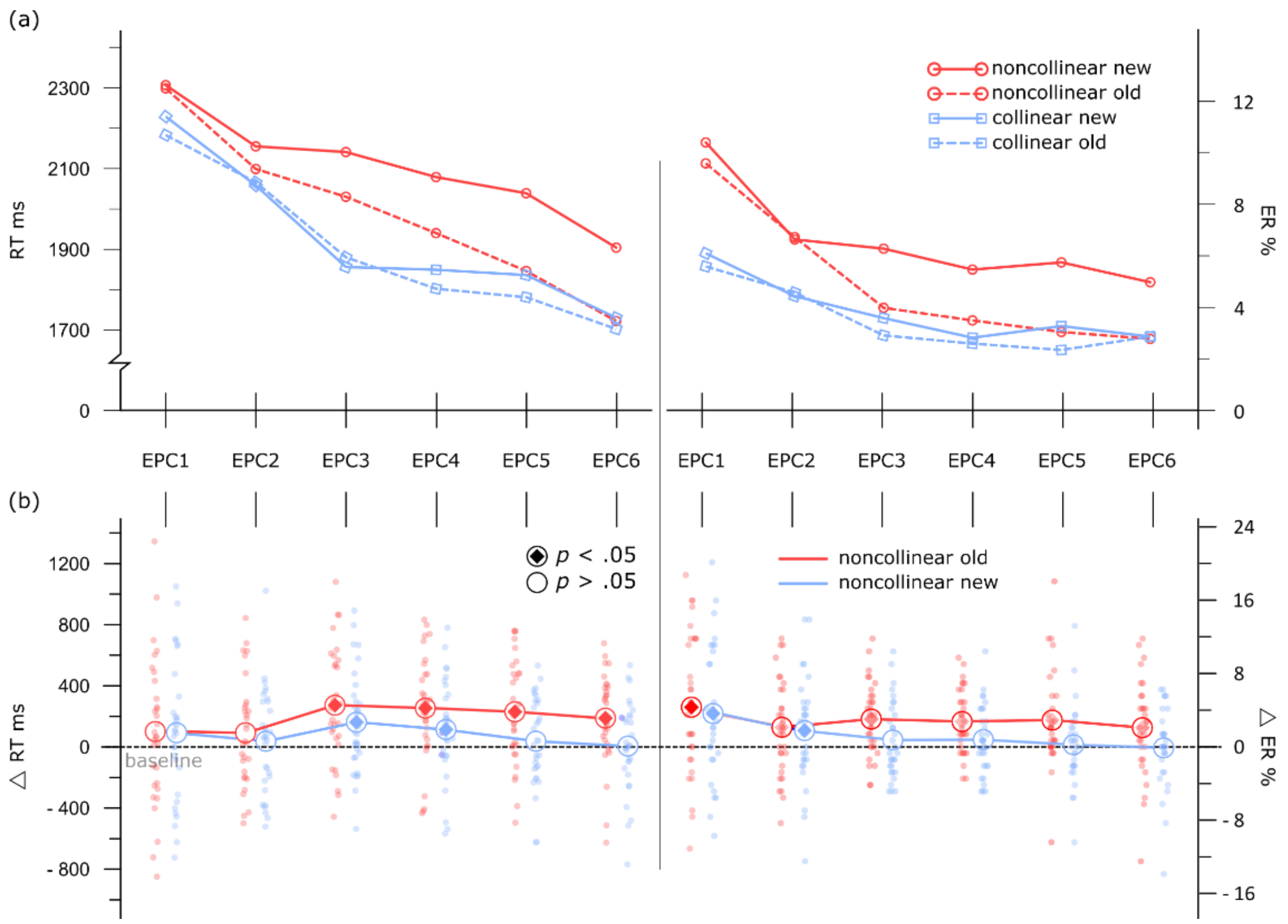

Figure 4. Results of Experiment 1. (a) Line charts display mean RT (left side) and ER (right side) for each layout group and each context type in each epoch. (b) The red lines display the mean RT (left side) and ER differences between noncollinear "new" trials and the baseline (the collinear group) in each epoch. The blue lines display the mean RT and ER difference between noncollinear "old" trials and the baseline in each epoch. The scattered dots represent the average RT or ER for each participant. A hollow circle indicates the difference was not significant; a circle filled with a diamond indicates the difference was significant.

Participants in both layout groups performed a yes/no recognition task. We applied one-sample *t*-tests to examine if participants' accuracy rate in the recognition was significantly better than random guessing ($M$ = 50%). We presented the 12 "old" screens randomly shuffled with 12 never-before-seen configurations and recorded both correct hits and false alarms. For the collinear group, the accuracy rate was 51.8%, which was not significantly better than random guessing, $p$ = .194. For the noncollinear group, the accuracy rate was 53.4%, which was not significantly better than random guessing, $p$ = .144. Participants in the noncollinear group also performed an additional locational recall task. The accuracy rate (10.1%) was not significantly better than random guessing ($M$ = 8.3%), $p$ = .125.

## 3.3 Discussion

The results of Experiment 1 accord with the hypotheses. Although participants in the noncollinear group initially showed longer RTs than participants in the collinear group, by Epochs 4 and 5 the RT in noncollinear old screen trials had caught up and were shown to be no slower than those of the collinear group. We attribute this to the CC effect. ER results showed a similar pattern (Figure 4). These results suggest that rather than searching for targets in every trial, when the noncollinear screens were repeated several times, participants started to recall the target locations. There was no such effect in the collinear condition. We hold that visual search was the primary means of target acquisition for the collinear arrangement, given the negligible difference in RT and ER between its "new" and "old" trials. In contrast, low-effort, automatic memory processes helped guide attention to the noncollinear target locations, making target accquisition for "old" trials as fast, and arguably more cognitively efficient, than when relying on primarily visual search. Most of today's digital device interfaces are arranged in fixed arrays similar to those in our collinear condition. Our results so far suggest that rectilinear alignment may deny users certain available efficiencies. The primary goal of Experiment 1 was to begin demonstrating the comparative effect on memory between a layout approximately similar to an actual human-computer interface and one spatially diverse alternative. However, the T/L stimuli presented in Experiment 1 are not representative of the actual working environment. Furthermore, while measuring speed may have some value for application, constantly resorting to speed at the cost of energy runs counter to a more practical goal, which should be to reduce effort in the long run. Speed may therefore become a secondary concern, and in fact CC's purpose is not to measure absolute speed but relative speed. The noncollinear arrangements showed both a distinct advantage in memorability, and the results also showed that with some small amount of training, rectilinear arrangements may in fact be no faster in absolute terms.

## 4 Experiment 2

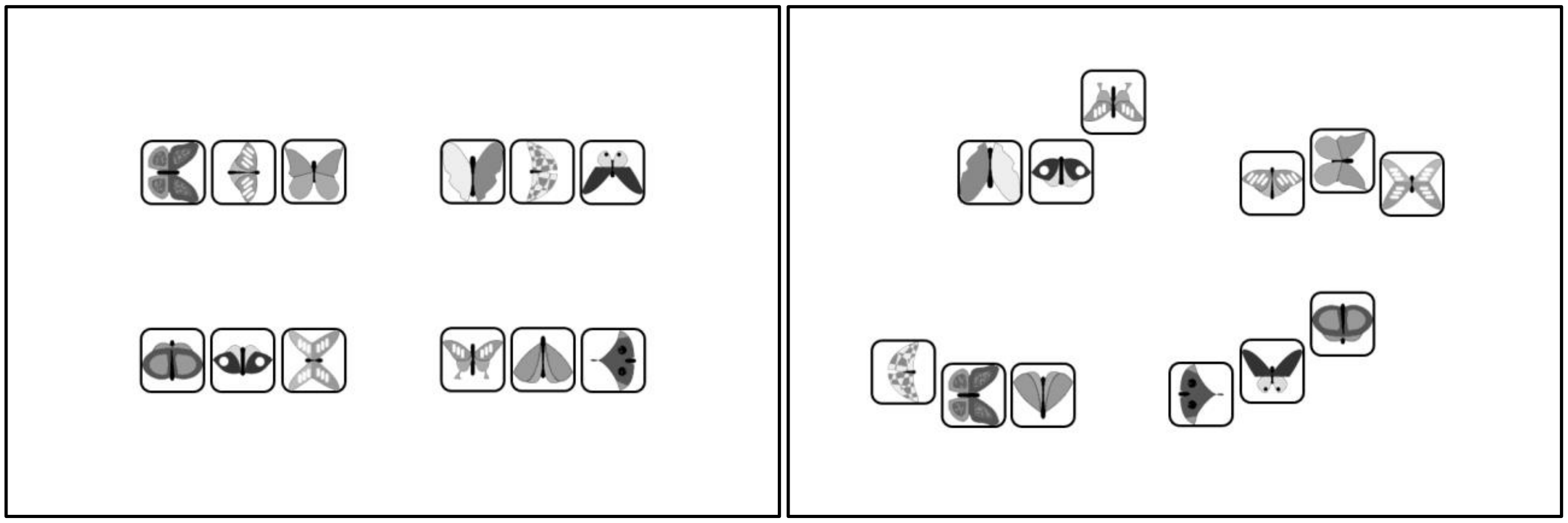

Figure 5. Representative displays used in the collinear (left) and noncollinear conditions of Experiment 2. These were taken from set A of the generated butterflies.

The stimuli used in Experiment 1 were the letters “T” and “L”, which differ markedly from targets and distractors (i.e., icons) used in human-computer interfaces. In order to further confirm the ecological validity of the experimental effects, here in Experiment 2 we set aside primitive psychophysical stimuli and began using butterfly icons as stimuli. Participants were required to locate the position of a specific grayscale butterfly in the experiment and report the direction of its flight among 11 distractor butterfly designs. Other than this, Experiment 2 was identical to Experiment 1.

We made the following predictions:

1. Participants in the collinear group should experience shorter RTs and lower ERs than participants in the noncollinear group.
2. Participants in the noncollinear group should experience a larger CC effect than those in the collinear group.

3. Due to the CC effect, after some training, in later epochs, the response delay in the noncollinear group should primarily exist in new screen trials, but the delay in old screen trials should be relatively small.

## 4.1 Method

### 4.1.1 **Design and Participants**

As with Experiment 1, in Experiment 2 we planned to investigate the contextual cueing effects in a three-way ANOVA with repeated measurements and within-between interactions. The results of a power analysis suggested an optimal sample size of $N = 62$ for $\alpha = .05$ and power = .80, with 31 participants in each layout group. A total of 66 participants from Hunan Normal University took part in Experiment 2. One student from the collinear group and one student from the noncollinear group did not achieve a sufficiently low error rate in each epoch (> 15%) and had to be replaced. A third student experienced a data storage error. Ultimately, we achieved a balanced sample of 37 participants in each layout group. Each participant received 50 RMB (≈ USD$8) for having taken part.

### 4.1.2 **Apparatus, Stimuli, and Task**

Apparatus for Experiment 2 was identical to that of Experiment 1.

#### *4.1.2.1 Stimuli generation*

Aside from internal features, array arrangement for Experiment 2 was identical to that used in Experiment 1 (see 3.1.3.1 above). This time, however, stimuli were to be more natural. We sought to identify a stimulus set that would approach the variety of feature detail and differentiation as that seen in actual daily use of interfaces, but that would neither carry intrusive semantic value nor otherwise be particularly memorable or nonmemorable to select viewers, as actual in-use or even generated tool controls and application icons might be. Initially, we had considered using photographs of dozens of species of butterflies; however, we found it difficult to adopt any systematic means of balancing their real-life features and to test them to confirm controlled conditions. Subsequently we settled upon artificial butterfly designs that were systematically generated based on a range of feature dimensions that varied with more distinctive differentiation than nature could provide for. The variations included color, size, wing shape, and special markings. After several phases of adjustments, we produced five butterfly sets of 12 stimuli each (see and Appendix). From among these, we selected three butterfly sets (A, C, D) deemed to have good configural variety and minimal anomalous features; from within each set we in turn chose a target that had only average or below-average feature prominence in its set (Appendix, Figure 16). We converted all stimuli to grayscale for this experiment.

##### ***4.1.2.2 Task***

The task was identical to that of Experiment 1, except for the different stimulus designs. The fixation cross was replaced with a larger image of the target butterfly, to remind participants of the target. While distractor butterflies could be rotated in any of the four primary compass directions, as with T/L and other experiments, the target could only point left or right so that the participant could easily respond with the “q” and “p” keyboard keys mentally mapped to left and right hands respectively. The main task was followed by the same recognition and recall tests as in Experiment 1.

### 4.1.3 **Procedure and Data Analyses**

Procedure and data analyses for Experiment 2 were identical to those of Experiment 1.

## 4.2 **Results**

Two identical three-way ANOVAs were conducted to compare the mean RTs and ERs between and within conditions. The two within-subjects factors were epoch (from Epoch 1 to Epoch 6) and context ("new" or "old" configuration). The between-subjects factors were layout group (collinear, noncollinear). Mean RTs, ERs, and corresponding *SD*s for each trial condition and participant group are shown in Figure X and listed in Appendix B, respectively. Results of the ANOVAs are summarized in Table X and illustrated in Figure 6.

-insert table X (or in appendix?)-

For the RT analyses, the main effect of epoch and main effect of context were significant. Participants' RTs decreased gradually from Epoch 1 (*M* = 1355 ms, *SD* = 384) to Epoch 6 (*M* = 1090 ms, *SD* = 238). Overall, participants tended to have shorter RTs in "old" screen trials (*M* = 1182 ms, SD = 304), than in "new" screen trials (*M* = 1210 ms, *SD* = 310). However, the RT difference between noncollinear group (*M* = 1207 ms, *SD* = 280), and collinear group (*M* = 1185 ms, *SD* = 323) across all six epochs was not significant. The interaction between layout group and context, and the interaction between epochs and context were significant. In order to better interpret the results, we first further tested the interaction between context and epochs in each layout group separately. The results indicated that the interaction between context and epochs was only significant in the noncollinear group, $F(5, 180) = 5.06$, $p < .001$, $\eta^2_p = .123$, but not significant in the collinear group, $F(5, 180) = 1.08$, $p = .371$. Next, pairwise comparisons showed that for the noncollinear group, the RT difference between "old" and "new" screen trials was significantly larger in the last three epochs (new – old = 58 ms) than in the first three epochs (new – old = 20 ms), $p < .001$, $d = +0.73$. For the collinear group, in the first three epochs, the RT difference between "old" and "new" screen trials was 13 ms, and it increased to 23 ms in the last three epochs. However, this increment was not significant, $p = .295$.

In Experiment 2, a significant CC effect was found for both the noncollinear group ("new" vs. "old" = 1158 vs. 1010, $p < .001$, $d = +0.25$) and the collinear group ("new" vs. "old"

= 1119 vs. 1095, $p$ = .009, $d$ = +0.08). Participants in the noncollinear condition experienced a larger CC effect (new – old = +58 ms) than participants in the collinear condition (new – old = +23 ms), $p$ = .004, $d$ = +0.68.

As with Experiment 1, for each epoch we compared the RTs of old screen trials and new screen trials from the noncollinear group with a baseline (the mean RT for the collinear group). The results were insignificant (all $p > .05$).

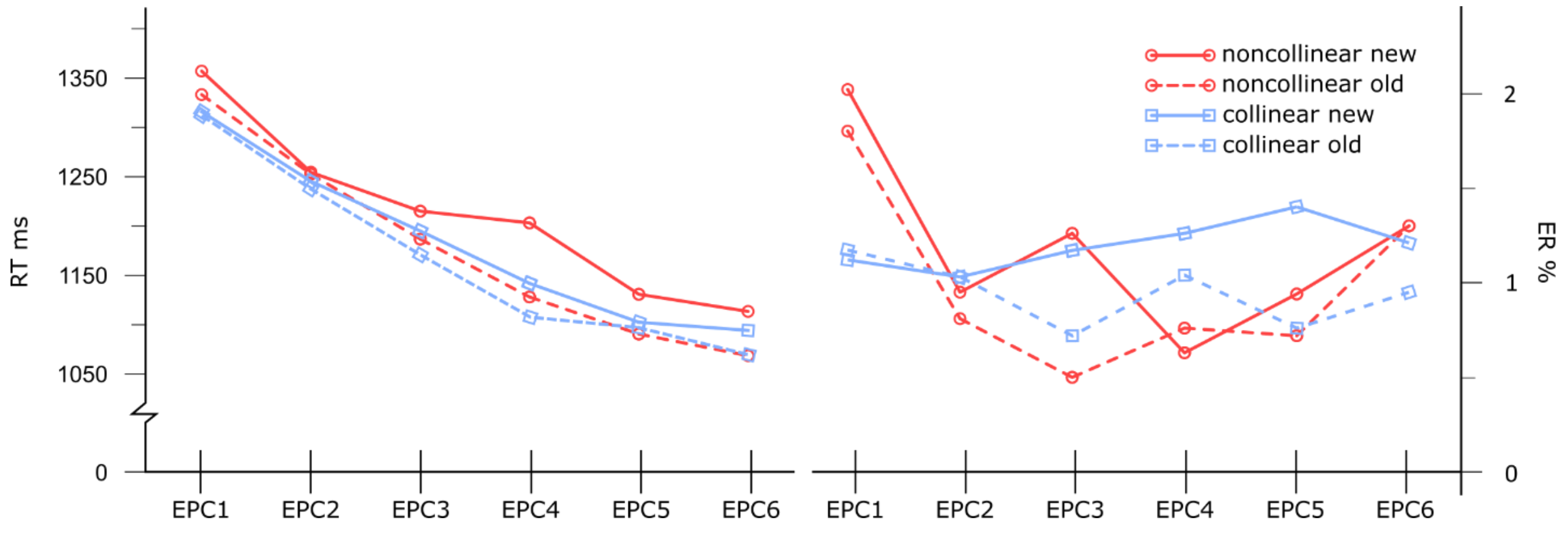

Figure 6. Results of Experiment 2. (a) Line charts display mean RT (left side) and ER (right side) for each layout group and each trial type in each epoch.

Participants in both layout groups performed a yes/no recognition task. We applied one-sample $t$-tests to examine if participants' accuracy rate in claiming recognition of the 12 "old" screens was significantly better than random guessing ($M$ = 50%). We presented the 12 "old" screens randomly shuffled with 12 never-before-seen configurations and recorded both correct hits and false alarms. For the collinear group, the accuracy rate was 49.6%, which was not significantly better than random guessing, $p$ = .855. For the noncollinear group, the accuracy rate was 57.4%, which was significantly better than random guessing, $p < .001$. Participants in the noncollinear group also performed an additional location recall task. Importantly, the accuracy rate (8.78%) was not significantly better than random guessing ($M$ = 8.33%), $p$ = .623.

For ER analysis,

## 4.3 **Discussion**

The sole difference between the designs of Experiments 1 and 2 was the change of target and distractor stimuli from T/L to more realistic artificial grayscale butterflies. Yet there were a number of differences in the results. Experiment 2 showed memory advantages for noncollinear by several measures: the stronger CC effect, as in Experiment 1; but also the marked difference between the two conditions in early- and later-epoch RTs; as well as recognition memory for the "old" configurations being significantly better than chance. We attribute these points to improvement in contextual learning, and we propose that this move from psychophysical to real stimuli offers a demonstration of a shadow of memory traces in transitional development from unconscious to recognition memories that simple target-distractor experiments may be unable to capture.

Participants recollected "old" configurations significantly better than chance. This is evidence of an incipient recognition memory. Here we find it puzzling that Experiment 1 saw both conditions approaching significance for recognition memory, and yet Experiment 2 shows its collinear condition's recognition memory figure lower than that of Experiment 1 and not nearing significance. This seems counterintuitive, since it would stand to reason that in Experiment 2 recognition for collinear would have improved along with that for noncollinear. The difference is not large, and so this may be a data artifact. However, we should offer an explanation under the assumption that participants were beginning to recognize "old" spatial configurations in both conditions and in both experiments. First, the fact that participants in noncollinear did better on recognition but failed to pinpoint target locations with any accuracy was likely due to noncollinear's greater spatial variations as contrasted with the physical and semantic interference among the butterfly features, as well as the relative global precedence of the spatial arrangements over the local features of the butterflies. As such, by the end of 30 exposures of each configuration, the spatial context will have become encoded as reliably as in Experiment 1, but specific target locations may have been further hampered both by the confusion of butterfly designs. This would stand to reason partly because the spatial variations are more global and in any case their differences are more pronounced for encoding and retrieval than differences among the stimuli. As to the objects' internal features, Experiment 1's "T" target, though deterred from preattentive parallel target acquisition by "L" distractors, is marginally distinct from the L's, which are all identical aside from rotation and reflection. On the other hand, while the distractor difference for butterflies appears substantial when compared among themselves, their designs consist of a greater distractor-distractor *difference* than the "L" patterns, and they also share greater target-distractor *similarity* in that there were greater numbers of perceptual as well as potentially differentiable Gestalt feature compounds (e.g., all butterflies had complex contours wings, bodies, antennae, etc.) and even feature and semantic saliency (e.g., I remember that butterfly with the spots) for comparison. It may stand to reason that this would result in intensified serial search when compared with simpler targets and distractors (Chun & Wolfe, 2005, pp. 279–283; Duncan & Humphreys, 1989; Wolfe, 1994, pp. 213–214). The fact that among the four conditions (two T/L + two butterflies) noncollinear butterflies served recognition memory best, and collinear butterflies worst, while the T/L arrays stood in the middle, must be explained by taking into account this difference in recall facilitation between noncollinear and collinear matter against the backdrop of object feature differences. This multiple-axis explanation may be clearly understood by visualizing the target and nontarget feature interrelationships against Duncan and Humphreys' (1989) search surface, and incorporating the role of relative spatial contextual similarity (Figure 7). Some might question the use of a model for effort to explain recognition memory. The CC paradigm is strongly influenced by the instance theory of automaticity (Logan, 1988, 2002). This theory holds that attention and memory are interdependent, and also that attentional effort influences memory encoding. Accordingly, we assume here that additional search effort will hinder the encoding processes to facilitate both recognition and location memory. The relative positions of the four CC experimental conditions in question along the search surface illustrate why the too-similar spatial context of collinear matter (both within each display *and* between displays) may interact with the lower distractor-distractor similarity among butterflies to reduce the total facilitative effect for memory. This suggests a dual concern when collinear matter is combined with more realistic object features than T/L in actual application: in the highly unpredictable competitive arena of tool, application, and object design, target designs will all compete for comparable saliency to their distractors. Presenting them all on the same spatial grid, as is commonly done, does nothing to allow for possibilities other than internal salient features for memory to latch onto.

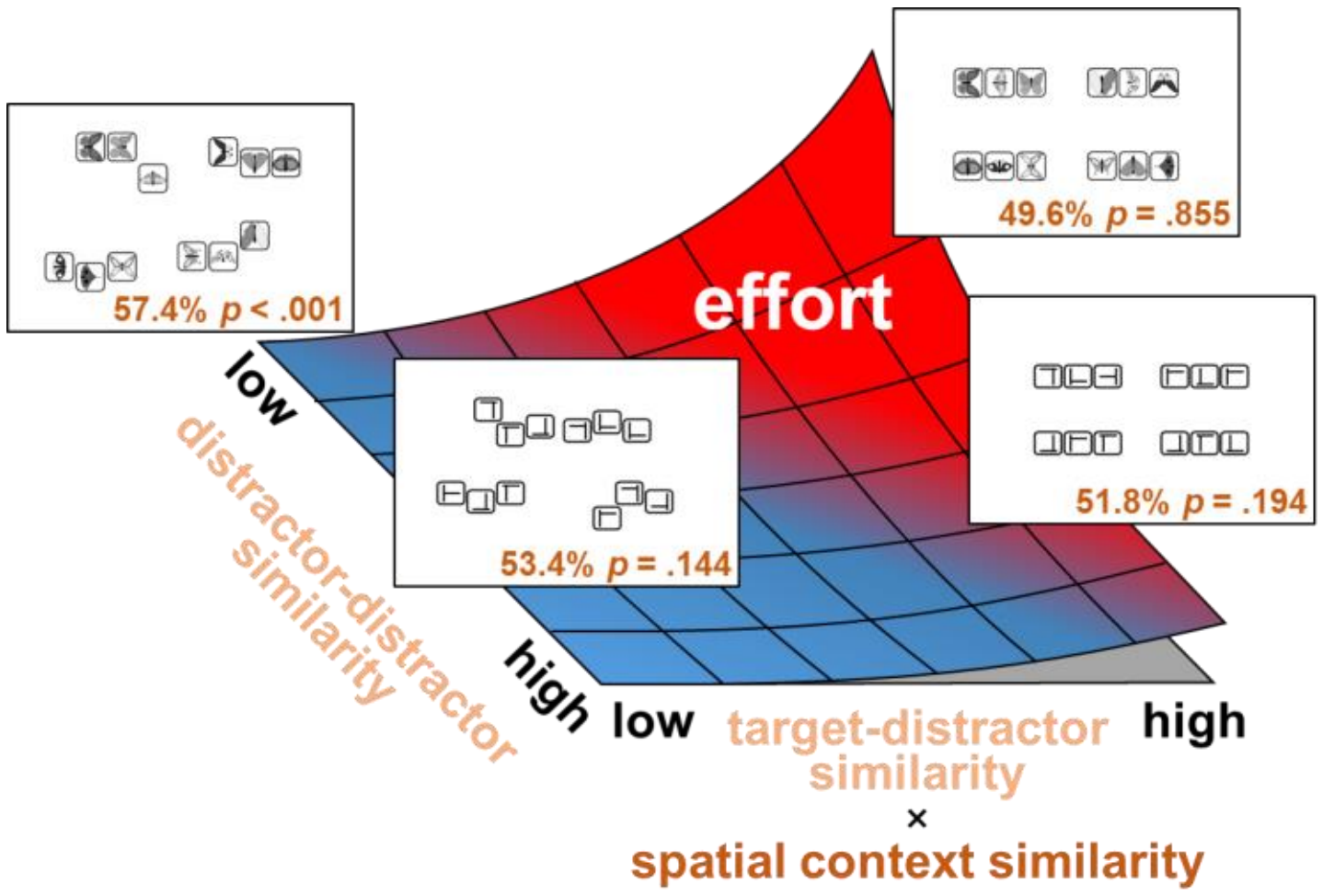

Figure 7. Explanatory illustration of recognition among the CC conditions using Duncan and Humphreys' target-distractor model, incorporating the role of spatial context similarity (search surface modified from Duncan & Humphreys, 1989, p. 442).

Despite the emergence of recognition memory for displays in the noncollinear condition, recall of exact target locations was still far below chance. If the above analysis is correct, then presumably this was still due to competition with "new" displays (Zang, 2014; Zang, Zinchenko, Jia, Assumpção, & Li, 2018; Zinchenko, Conci, Müller, & Geyer, 2018). Now that we could see incipient traces of memory under increasingly natural object conditions, we felt it would be appropriate to move further up the chain of custody in naturalism to understand more about how visual search and recognition memory are developed or inhibited, by using global spatial variation in a more realistic task.

# 5 Experiment 3

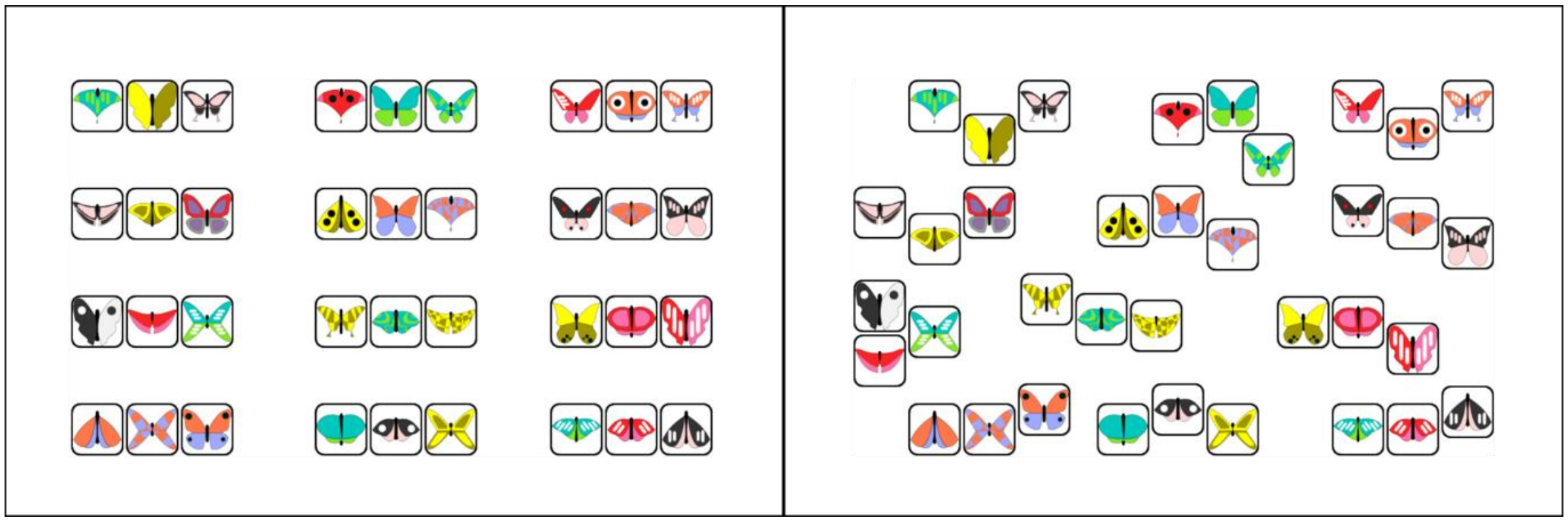

Figure 8. Representative displays used in the collinear (left) and noncollinear conditions of Experiment 3.

At this point in our investigation, it was clear that the contextual cueing evidence was sufficient to confirm our hypothesis of a basic difference in search efficiency for collinear and noncollinear arrays. But, does this hold when we diverge from psychophysical evidence and further broach HCI? And, will users in more practical environments do better and consciously recall more? The goal was to examine the difference between collinear and noncollinear layouts in a more realistic HCI scenario. Specifically, Experiment 3 required participants to click target icons using a mouse, in a search activity similar to those of the CC environment except using a full-screen display search for multiple colored targets. Importantly, we only showed one display arrangement per condition, and prompted for a larger number of targets to be sought among a larger number of distractors. This generic task simulates many kinds of interactive activities on digital devices. The experimental benefit of the single-screen approach is that without the interference from “new” screens, development of recognition memory traces should be more rapid (Zang, 2014; Zang et al., 2018; Zinchenko, Conci, Töllner, Müller, & Geyer, 2020). Because of this, we should be able to begin to see greater evidence of recognition memory and finally location recallability, as well as compare errors. The task falls into the category of a quasi search game or speed exercise for participants, marginally more engaging than the CC experiments.

We were particularly interested in understanding more about collinear location recall errors. Anecdotally, it is rather clear that collinear arrays frequently cause visual confusion (e.g., “which drawer were the spoons in?”). If scene instances are automatically encoded and automatically retrieved, then it follows that encoded collinear arrays should cause greater location confoundment with collinear nearby objects. We believe that this may be built into modern GUI use and may substantially increase cognitive load. Therefore, it should be of value both to theory and practice. However, despite careful search and inquiry, we have seen nothing in cognitive or HCI literature that directly discusses this.

Finally, we were interested in the specific nature of the errors. If an object’s location is incorrectly remembered, then a user’s preformed goal for reaching for the object in the wrong location will influence target acquisition efficiency in the future. Drawing from Tversky and Schiano (Tversky, 1981, 1990; Tversky & Schiano, 1989), as well as various other studies about spatial relations in topographic views (Rossano & Morrison, 1996; Rossano & Warren, 1989), we hypothesized that rectilinear arrangement may increase recall errors, particularly vertically and horizontally collinear errors.

We made the following predictions:

1. There should be no difference in terms of RT and ER between the noncollinear and collinear groups.
2. Participants in the noncollinear condition should have better performance in the item recall tasks on the first day and on subsequent days.
3. Participants in the noncollinear condition should experience lower collinear relocation confoundment, whereas participants in the collinear condition should experience higher collinear relocation confoundment, mislocating more items to near neighbors collinearly vertical and horizontal to the actual target.

## 5.1 Method

### 5.1.1 **Design and Participants**

Experiment 2 had two parts. In the first part, we compared participants' mean RT and ER in a two-way factorial experimental design. The between-subjects factor was the layout group (noncollinear vs. collinear), and the within-subjects factor was epoch (1, 2, or 3). Based on our prior predictions, we reasoned that if a difference in recall between noncollinear and collinear groups does exist, at least initially it would be very small. Hence, in the statistical power analysis (Faul et al., 2007), we expected to have a small effect size ($f$ = 0.15). The results of the analysis suggested an optimal sample size of $N$ = 62 for $\alpha$ = .05 and power = .80, with 32 participants in each layout group. In the second part, we would compare the mean recall in a two-way factorial experimental design. Two factors were layout (noncollinear, collinear), and location recall accuracy (Day + 0, Day + 1, Day + 3, and Day + 7). The statistical power analysis (Faul et al., 2007) suggested an optimal simple size of $N = 24$ for $\alpha = .05$, $f = .25$ and power = .80, with 16 participants in each layout group. Taking both power analyses into consideration, we decided to collect a sample of $N = 62$, with 31 participants in each group. Ultimately, 66 participants participated in Experiment 3, with 33 participants in each group.

### 5.1.2 Apparatus, Stimuli, and Task

Experiment 2 was programmed using the PsyToolkit online platform (Stoet, 2010, 2017). An arrangement of 36 butterflies appeared on screen, arranged in 4 rows and 3 columns of triplets in the collinear condition, closely resembling the layout differences in the CC experiments. As with the CC experiments, the noncollinear condition would feature a variety of spatial configurations based on the same considerations of spatial entropy. We took care to obtain a natural optical spacing between items and hierarchical groups both within and between the two conditions. However, we maintained some alignment to provide for the possibility of detecting collinear relocation errors in both conditions. Stimuli were taken from the same butterfly designs used in Experiment 2, except that in this case we used their full colors. The task for each participant was to locate each of 12 different butterflies in the search array, when randomly prompted with one butterfly design. Using butterfly sets A, B, and C, to control for effects of stimuli, we balanced the designs across conditions by rotating each design set of 12 (e.g., A) in as the target set for the given condition, with the remaining two sets of 12 (e.g., B and C) serving as the distractors for that group.

The main search activity was presented in nine blocks of 24 trials each, with an optional one-minute rest permitted between each block. In each block of 24 trials, the 12 target butterflies were each presented, twice, in random order. After a fixation cross, a larger version of the prompt target would appear briefly, followed by the search screen, which as we have mentioned does not change throughout the experiment. The participant was permitted up to 5 seconds to locate and click the mouse on the target, after which correct or incorrect feedback would be shown. Then, the next randomly selected butterfly of the 12 targets would be presented for search. Therefore, each target would be searched 18 times during the full experiment of 9 blocks × 2 times per target per block. The same task was given in a practice block before beginning the actual experiment. A trial scheme including the time sequence is provided in Figure 9.

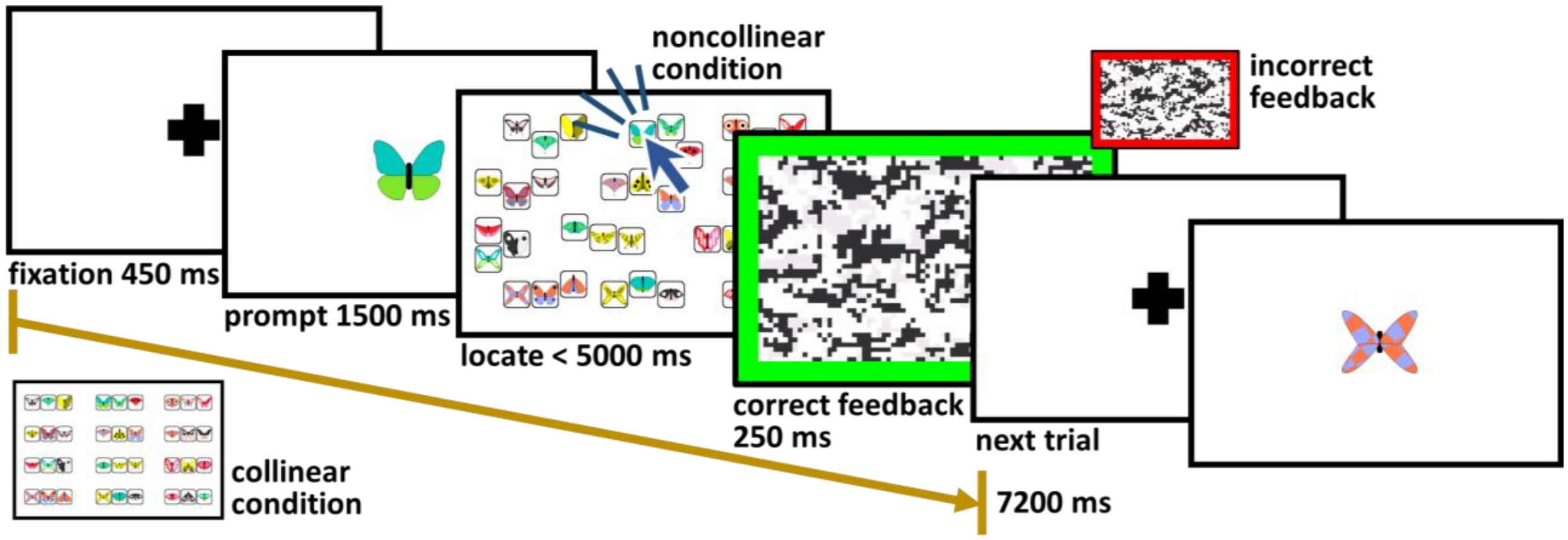

Figure 9. Trial scheme for Experiment 3.

### 5.1.3 **Procedure**

Students from Fudan University first signed up with a research assistant, who randomly assigned each participant to the collinear layout group or the noncollinear layout group. Participants were seated alone in a room selected to limit distractions. Once seated, participants were asked to click a box to confirm that they had agreed to participate in the experiment session and informed that they had the right to withdraw at any point. After that, participants read on-screen instructions and then completed the tasks in each sesson. Aside from the main session, there were three other experimental sessions, brief in duration. Participants in the collinear layout group completed the collinear version, and participants in the noncollinear layout group completed the noncollinear version.

In the main session (Day + 0), participants were first asked to complete the butterfly searching task; following this, a screen full of random digits appeared, and participants were asked to count the number of 3's on the screen for 30 seconds as a working-memory clearing activity. They then completed a surprise recall and confidence task. On the second (Day + 1), third (Day + 3), and fourth contacts (Day + 7), participants were asked to complete the same recall and confidence task again. After having finished all four contact sessions, the research assistant transfered 50 RMB to the participant via Alipay.

#### 5.1.4 **Data Analyses**

For the butterfly searching task, we aggregated three blocks into one epoch (e.g., blocks 1–3 = Epoch 1). Participants had to complete 9 visual searching blocks, hence we aggregated them into three epochs. Incorrect trials we excluded from RT analyses.

We also needed to evaluate our hypotheses for collinear relocation memory errors. After recording each specific error for each condition during the training-day posttest and on each of the three subsequent days during the week, we planned to analyze patterns of error-making spatially, including how the types of errors developed with the passage of time. We made an assumption that errors approximately within the attentional spotlight were potentially relevant to local spatial confusion and location confoundment, and we suspected that errors would often be made to collinear vertical and horizontal nearest neighbors. The spotlight of attention for various tasks is difficult to pin down. However, CC literature studying inter-object effects suggested varying areas of interest for spatial contextual influence of between 12° and 18° of visual angle (Conci et al., 2013; Kimchi et al., 2016). We decided on a 10°-diameter circle as a standard for Experiments 3 and 4, as it satisfied some modicum of operationalization for a spotlight under some time stress; it limited the area of interest to near neighboring objects and hierarchical groups; and it also allowed for the inclusion of the nearest object in the neighboring horizontal group. We settled on the criterion that at least one half the width of the edge of an object's rectangle must fall within the circle to be considered a within-circle object.

We computed the Euclidean distance between each target's center and the center of the distractors, including in the analysis only distractors which fell within the 10° circle around the target. Of these, we counted collinear vertical and horizontal errors, as well as errors within the set. Since the collinear condition would by its design accrue more such collinearity errors than the noncollinear condition, we determined the probability of each error around each target in each condition based on the number of possible errors of that type within the circle. This would level the playing field for the two conditions. We planned a binomial test for each of the 12 targets × type of error (within set, collinear vertical, collinear horizontal) × day (Day + 0, Day + 1, Day + 3, Day + 7). Finally, we also developed a simple visualization tool to show where the within-circle errors occurred.

## 5.2 Results

### 5.2.1 RT and ER

Two identical two-way ANOVAs were conducted to compare the mean RTs and ERs between and within conditions. The within-subjects factors were epochs (Epochs 1, 2, and 3). The between-subjects factors were layout group (collinear, noncollinear). Mean RTs, ERs, and corresponding *SD*s for each trial condition and participant group are shown in Figure 10 and listed in Appendix B, respectively.

For RT analysis, the main effect of epoch was significant, $F(2, 128) = 471.23$, $p < .001$, $\eta^2_p = .880$. Participants' RT reduced gradually from Epoch 1 to Epoch 3. The main effect of layout was not significant, $F(1, 64) = 0.05$, $p = .821$. The interaction between two main effects was not significant, $F(1, 64) = 2.43$, $p = .091$.

For ER analysis, the main effect of epoch was significant, $F(2, 128) = 109.76$, $p < .001$, $\eta^2_p = .632$. The main effect of layout was not significant, $F(1, 64) = 1.33$, $p = .252$. The interaction between two main effects was significant, $F(2, 128) = 10.04$, $p <.001$, $\eta^2_p = .135$. Participants' ER reduced gradually from Epoch 1 to Epoch 3. Moreover, the interaction between epoch and layout group was also significant. Pairwise comparison showed that in Epoch 1, participants in the collinear group (13.89%) had higher ERs than than participants in the noncollinear group (9.39%), $p = .024$, $d = +0.27$. In Epoch 2 (collinear vs. noncollinear = 5.43% vs. 4.71%, $p = .641$) and Epoch 3 (collinear vs. noncollinear = 2.03% vs. 3.07%, $p = .280$), the ER difference between two layout groups was not significant.

### 5.2.2 **Recall**

#### ***5.2.2.1 General differences***

A two-way mixed ANOVA was conducted to compare the mean recalled butterfly locations between and within conditions. The within-subjects factor was recall day (Day + 0, Day + 1, Day + 3, Day + 7), and the between-subjects factor was the layout group (noncollinear, collinear). The results showed that the main effect of layout group was significant, $F(1, 64) = 6.78$, $p = .011$, $\eta^2_p = .096$. The main effect of recall day was significant, $F(3, 192) = 89.09$, $p <.001$, $\eta^2_p = .581$. The interaction between two factors was also significant, $F(3, 192) = 5.20$, $p = .002$, $\eta^2_p = .075$. The pairwise comparison suggested that on Day + 0, participants in the noncollinear group (10.30) and collinear group (10.21) recalled similar numbers of butterfly locations, $p = .837$. On Day + 1 (noncollinear vs. collinear = 9.03 vs. 7.42, $p = .034$, $d = +0.66$) and Day + 3 (noncollinear vs. collinear = 8.12 vs. 6.09, $p = .009$, $d = +0.75$), participants in the noncollinear group remembered significantly more butterfly locations than participants in the collinear group. On Day + 7, participants in the noncollinear group (6.96) also remembered more butterfly locations than participants in the collinear group (5.48), but the difference was not significant, $p$ = .053.

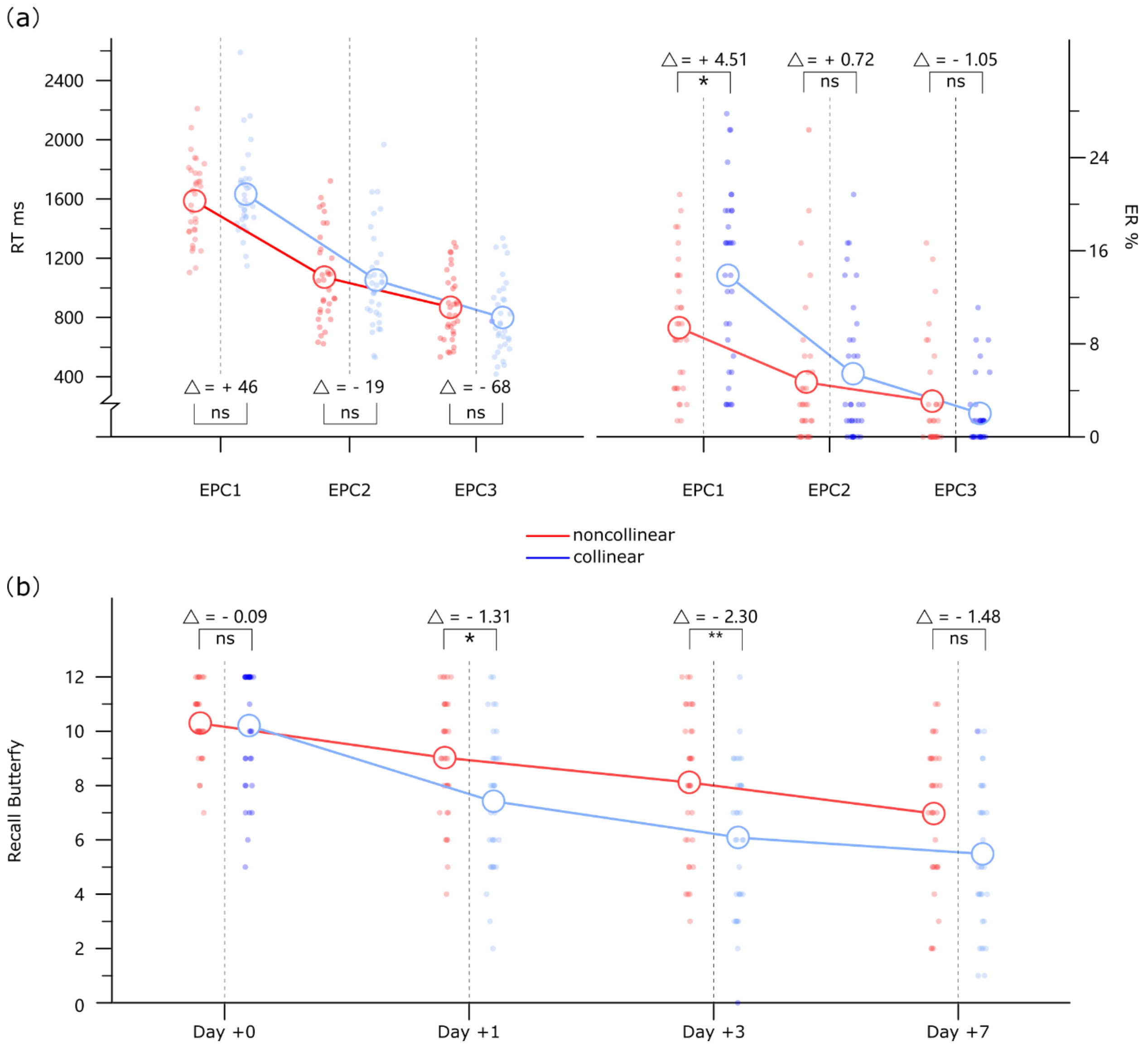

Figure 10. Results of Experiment 3. For all charts, red indicates noncollinear and blue indicates collinear. (a) Line charts display mean RT (left side) and ER (right side) for each layout group and each epoch. (b) The red lines display the mean RT (left side) and ER differences between noncollinear "new" trials and the baseline (the collinear group) in each epoch.

#### *5.2.2.2 Specific location error patterns*

Binomial tests suggested substantial differences in the nature of errors between the two conditions. We analyzed errors within the defined circle (see 5.1.4 above) to be within the same set (WS), collinear vertical (CV), collinear horizontal (CH), and/or other location within the circle (OT). A visualization showing relocation error progress for the two conditions on each day and for each target can be seen in Appendix X. A table including hit probabilities, binomial test *p* values for each location on each day, and specific error counts for each error category (within-set, collinear vertical, and collinear horizontal) for each location on each day can be found in Appendix X. Note that some errors were necessarily double-classified as WS and CH, and we computed probabilities and statistics for these separately. Also, we did not classify confoundments with targets within the circle but in different sets, but this could have had an influence on memory and in some cases it is difficult to ascertain whether these are due to misidentification of the set or of the target spatial location.

For both conditions, errors were quite few immediately after training. Errors, including errors outside the circle, increased over days for both conditions. However, the increases were greater for the collinear than for the noncollinear condition.

Several collinear vertical (CV) errors predominated on the inside rows (rows 2 and 3) of the display array, and these began coming to significance on Day + 1 and continued to be significant through Day + 7. These CV errors were most prominent on the third of the four rows, and subjects most frequently erred upward to immedately collinear vertical distractors in the second row. As errors from clusters at the top and bottom of the array, CV errors were uncommon. However, collinear horizontal (CH) errors were found at levels significantly above chance primarily in the perimeter rows (1 and 4) of the collinear condition. Note that particularly in collinear condition, errors within the set of three (WS) appeared primarily in the perimeter clusters, and CV errors were fewer.

The binomial tests also showed some apparent patterns among relocation errors in the noncollinear arrangement in particular. Interestingly, collinear vertical errors were observed in the *noncollinear* condition as well. However, noncollinear saw more erroneous relocations in the same position in neighboring clusters on the diagonal.

Both CV and CH errors in the collinear condition more than doubled between Day + 0 and Day + 1 (CV = 17 to 42, CH = 20 to 56). Interestingly, CV and CH errors in the noncollinear condition nearly tripled between Day + 0 and Day + 1 (CV = 6 to 29, CH = 7 to 19). The latter are relevant because they are collinear vertical and horizontal errors despite

By Day + 7, errors outside the circle in both conditions had become more common as participants misplaced more and more targets.

## 5.3 Discussion

Errors in the collinear condition showed persistent patterns of mistaken relocation to immediately vertical neighbors, and both conditions showed confusion with horizontal neighbors, although due to the experiment design, the evidence that the latter is due to collinearity is less clear than in the case of the vertical errors. This experiment provided some initial evidence that, contrary to widespread belief among GUI researchers and practitioners, collinearly arranged layouts may not serve visual search speed any better than noncollinear layouts. Of greater concern is that the collinearity may in fact induce vertical and horizontal errors, whereas more freely arranged displays would decrease the chance of this kind of error. This all may seem counterintuitive.

We should note here an issue with CC that extends to real environments and perhaps helps to explain a phenomenon of concern to HCI. With the CC experiments, layouts of 360 “new” and 12 “old” collinear arrays were all spatially very similar. Even in the noncollinear condition, each of the 360 “new” and 12 “old” arrays, though distinct, was distinctive due to the exact same geometric formula. Like 101 dalmatians, the spots had different qualities and appeared in different places on each body, but otherwise the bodies were essentially all the same. Given time, however, the mother dog begins recognizing certain salient offspring. Naturally, this problem of distinctiveness did not exist in our single-screen activity, and it ought not to exist in scene environments of one or only a few scenes of similar design, such as a simple single-screen app or the rooms of a home. However, in collinear matter, environments may be laid out identically along numerous instances, such as with hotel rooms or supermarket shelves or screenfuls of data, or even similarly but differentially along numerous instances, such as with visualizations or phone preference settings. When this happens, the similar instances, all on the same figural plane, seem to occlude one another.

A careful look at the nature of errors in the collinear displays suggests some apparent tendencies to make more collinear vertical relocation errors when the target lies inside the periphery; to prefer collinear horizontal relocation errors when the target lies along a perimeter; and in general to err inward toward the center of the global space rather than outward. Given only four rows, it is not clear whether vertical collinear errors were predominantly downward or vertically toward the center. In any case, the closest theoretical explanation for this phenomenon may be from studies on topographical map interpretation by Rossano and Morrison (1996), who found a “peripheral learning bias” (pp. 115–116) wherein locations near an edge were better recalled. Treating peripheries *per se* as a special sclass of landmark (Siegel & White, 1975) is one way that helps us to explain this, as the edge of a scene would certainly qualify as a prominent structural feature, a powerful anchor for spatial reference. Getting lost inside a structure seems to be quite common, and made more difficult by collinearity.

# 6 Experiment 4

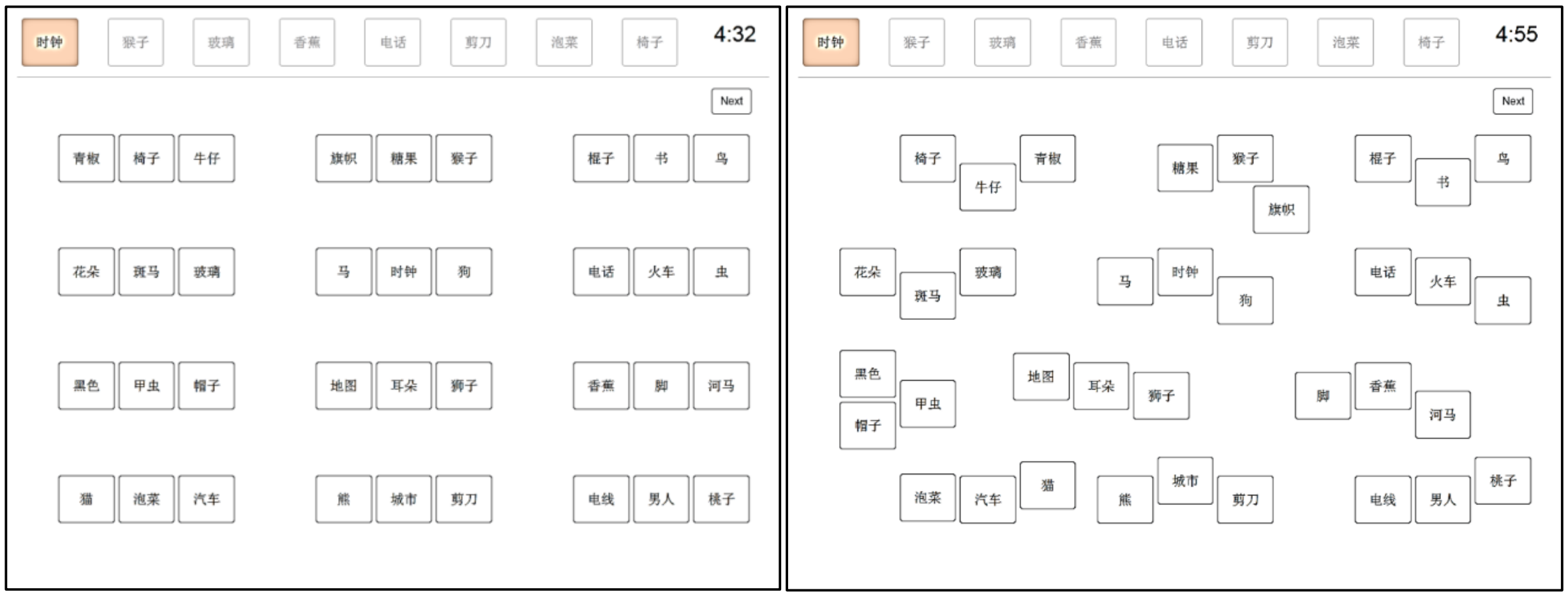

Figure 11. Representative displays used in the collinear (left)
and noncollinear conditions of Experiment 4.

The goal for Experiment 4 was to examine the difference between noncollinear and collinear layouts in a scenario yet more environmentally faithful to actual human-computer activity. Throughout this series of experiments, we attempted to hold the global and local visuospatial qualities fairly consistent, but in this case we wanted to the user to become absorbed in an activity that would cause them to ignore the interface as an object of interest. An area of continual interest in HCI study is group support systems (GSS). One of the most common GSS applications is electronic brainstorming (EBS). One concern with EBS involves distinguishing cognitive task load attributable respectively to the interface and to the task (Sweller et al., 1998). However, purely cognitive experiments are criticized for lacking realism and therefore applicability (Jung, Schneider, & Valacich, 2010; Cyr, Hassanein, Head, & Ivanov, 2007). Therefore, we opted to create a simplified individual EBS divergence task to test recall and cognitive load, holding other aspects constant apart from the conditions and counterbalances.

We also wanted to see how comparatively well individuals would remember locations given a much briefer exposure. The activity would involve words instead of abstract shapes or pictures. Participants would seek and click on target words, as they did with butterflies in Experiment 2, but this would not be a rapid search activity. The difference would be that in this activity they would look for each target only once. As with Experiment 2, we also included only one screen in the entire experiment, but we would keep the activity short. However, unlike Experiment 2, given the short duration of training we expected lower levels of incidental recall.

We made the following predictions:

1. Participants in the noncollinear condition should demonstrate better object location recall in a surprise target location recall task.
2. Perceived task load should be comparable in both conditions.
3. Confoundment of correct target locations with nearby collinear distractor locations would occur frequently in both collinear condition and in noncollinear condition where there were opportunities to do so.

## 6.1 Method

### 6.1.1 Research ethics disclosure

All participants were informed in advance of the nature of the experiment and registered informed consent by signing a form.

### 6.1.2 Design and Participants

As with the previous experiment, we chose two screen arrangements, collinear and noncollinear, with 36 items arranged in the same pattern. Exemplars of the two experimental displays are shown in Figure 11. The experiment consisted of three parts. In the first part, participants worked through a brief set of tasks in an individual brainstorming simulation. The second portion was a surprise location recall test, finally followed by a cognitive load evaluation using the NASA/TLX. The brainstorming portion was timed to be completed within 5 minutes and the remainder would take only three to four additional minutes.

A total of 96 undergraduate students (48 male, 48 female, age $M$ = 19.8 years, $SD$ = 1.2 years) from the University of Michigan-Shanghai Jiao Tong University Joint Institute participated in the study in exchange for course credit. We determined that we would require 48 participants per condition for sufficient statistical power, and we would also split these groups for balanced conditions in the control check on semantics.

### 6.1.3 Apparatus, Stimuli, and Task

We used a bank of 24 Dell workstations reserved in a secure, quiet study room at the University of Michigan-Shanghai Jiao Tong University Joint Institute. The workstations were equipped with 24-inch Dell monitors with 1920 × 1080 resolution and set to medium brightness and distance. Students were briefed and debriefed in several groups.

A row of 8 simple target topic words in Chinese was shown at the top of the screen. An arrangement of 36 words, including all of the words from the top, appeared at the bottom, replacing the butterflies of Experiment 2.

Words were selected from two separate corpora. We took 8 words from the multilingual list of 10 words in the Common Objects Memory Test (COMT) (Kempler et al., 2009). Another 16 words came from Pezdek and colleagues (1986). We evaluated a paper/PDF pilot in English with 19 international graduate students. With this lexicon, some subjects felt the activity was not engaging enough. After discussion with subjects, we incorporated 5 unusual words (pickle, glass, monkey, phone, banana) to spark greater interest. Finally, for the 12 distractors, we used words from the same published corpora, as well as 3 additional words with reasonable cultural neutrality. Words not in the COMT were translated into Chinese and re-translated using separate native speakers. This list is shown in **Error! Reference source not found.**. Each word on the screen appeared in a 2 cm-square outline box in a sans serif font (14-pixel SimSun Chinese for display; 18-point Calibri English for the paper pilot). Each topic target word was grouped with two non-target partner words in the region below the stimulus row, with a space of 4 pixels (3 mm) between group items. In a pilot, the activity array contained only the 8 target stimuli items plus their 2 partners each, totaling 24 items. As with Experiment 3, we ultimately used 36 items in 12 sets of 3 items. However, to ensure some consistency between the two conditions, we arranged the 8 target topic items in the same absolute global on-screen positions in conditions, similar to other location recall experiments (Kulhavy, Schwartz, & Shaha, 1982; Rossano & Morrison, 1996). In this way, the absolute location would be in the same location on screen for both conditions. Finally, to control for linguistic or semantic interference, we shifted the entire word list for half of the subjects to different sequential positions.

Participants were instructed to go through 8 target stimulus nouns (in Mandarin; see Table 2) highlighted one by one at the top of the screen. The task was first to locate the highlighted stimulus word shown above from the screen of 36 icons below, then to observe the two remaining stimuli in that group of three and devise an amusing sentence using those three nouns without writing it down. For example, Target t0, 时钟 “clock,” was found in the middle of the second row, with 马 “horse” and 狗 “dog.” The participant might take a few seconds or longer to think of a sentence like, 狗把马的始终偷走 “The dog stole on the horse’s clock.” They would next proceed with Target t1, and continue until they had processed all 8 targets. Time was kept during this part. After this, a 2-minute clearing task to count 3’s among random digits was given (Pezdek, Roman, & Sobolik, 1986b; Pirolli, Card, & Van der Wege, 2001).

Following this, a surprise object relocation task was given for each of the 8 targets. Finally, participants completed a paper version of the NASA/TLX (Hart & Staveland, 1988) translated to Mandarin (Xiao (肖元梅), 2005). Participants were then debriefed.

Table 2. Experiment 4, Brainstorm Item Words in Chinese and English

| | | |
|---|---|---|
| 青椒 牛仔 椅子<br>pepper/cowboy/**chair (t7)** | 猴子 糖果 旗帜<br>**monkey (t1)**/candy/flag | 棍子 书 鸟<br>stick/book/bird |
| 玻璃 斑马 花朵<br>**glass (t2)**/zebra/flower | 马 狗 时钟<br>horse/dog/**clock (t0)** | 火车 电话 虫<br>train/**phone (t4)**/bug |
| 黑色 甲虫 帽子<br>black/beetle/hat | 地图 耳朵 狮子<br>map/ear/lion | 河马 脚 香蕉<br>hippo/foot/**banana (t3)** |
| 猫 泡菜 汽车<br>cat/**pickle (t6)**/car | 熊 剪刀 城市<br>bear/**scissors (t5)**/city | 电线 男人 桃子<br>wire/man/peach |

Note: Topic words in **bold**. Groups with no bold item contain all distractors.

## 6.2 Results

Relocation counts per target item are shown in Table 3, and a comparison between the two conditions on correct-to-target locations by participants is shown in a bar chart in Figure 12. Among both conditions, participants correctly identified the target location an average of 1.68 times out of 8 possible (*SD* = 1.55).[2] Participants in the collinear condition remembered fewer total targets on average than in noncollinear (*M* = 1.2 vs. 2.2; see also Figure 12). Furthermore, fewer than one in three collinear participants correctly identified locations of 2 or more targets, while nearly two-thirds of noncollinear participants did so.

Table 3. Experiment 4, Correct Relocations per Target Item

| | **t0** | **t1** | **t2** | **t3** | **t4** | **t5** | **t6** | **t7** | **Correct *M*** | **Correct *SD*** |
|---|---|---|---|---|---|---|---|---|---|---|
| **All (96)** | 59 | 19 | 29 | 9 | 16 | 11 | 8 | 10 | 20.1† | |
| **Collinear (48)** | 25 | 5 | 8 | 4 | 5 | 4 | 2 | 4 | 1.2 | 1.2‡ |
| **Noncollinear (48)** | 34 | 14 | 21 | 5 | 11 | 7 | 6 | 6 | 2.2 | 1.7‡ |

†Of columns, mean correct relocations per target. ‡Of rows, average correct responses per item.

[2] A primacy effect, consistent across all four conditions, was apparent for the first and some subsequent items.

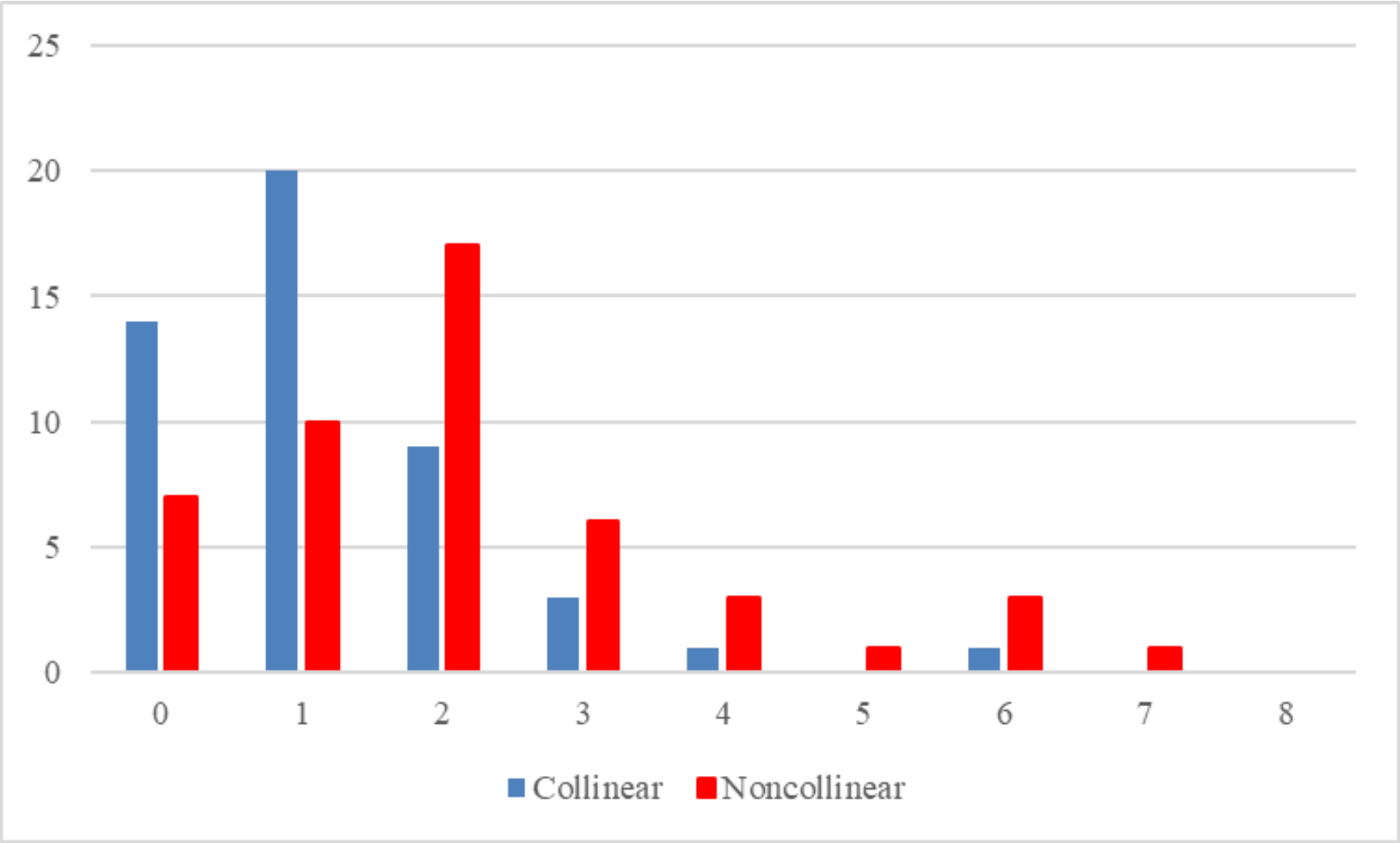

Figure 12. Experiment 4, Number of Item Locations Correct per Participant

### 6.2.1 Specific location error patterns

As with Experiment 3, we examined in greater detail the nature of relocation errors within a 10° circle from each target. A summary of error patterns is shown in Table 4. In-circle errors were approximately the same for both conditions (*M* = 2.60 vs. 2.33, *p* = .332). A difference in number of errors out of the circle, while higher on average for collinear (*M* = 4.21) than noncollinear (*M* = 3.50), approached but did not reach significance (*p* = .056).

Table 4. Experiment 4, Summary of Errors

| | | **Overall** | | **Collinear** | | **Noncollinear** | | |
|---|---|---|---|---|---|---|---|---|
| | | ***M*** | ***SD*** | ***M*** | ***SD*** | ***M*** | ***SD*** | ***p*** |
| **Correct to target** | † | 1.68 | 1.55 | 1.19 | 1.20 | 2.17 | 1.71 | .002** |
| **In-circle errors** | † | 2.47 | 1.36 | 2.60 | 1.23 | 2.33 | 1.48 | .332 |
| **Out-of-circle errors** | † | 3.85 | 1.82 | 4.21 | 1.71 | 3.50 | 1.87 | .056 |
| **Within-set errors** | | 0.79 | 0.82 | 0.83 | 0.91 | 0.75 | 0.73 | .621 |

†Column means add to 8 trials per subject. ***p* < .01.

Binomial tests suggested differences in the nature of errors between the two conditions. We analyzed errors within the defined circle (see 5.1.4 above) to be within the same set (WS), collinear vertical (CV), collinear horizontal (CH), and/or other location within the circle (OT). A visualization showing relocation error progress for the two conditions on each day and for each target can be seen in Appendix X. A table including hit probabilities, binomial test *p* values for each location on each day, and specific error counts for each error category (within-set, collinear vertical, and collinear horizontal) for each location on each day can be found in Table 5.

The results showed vertical and horizontal errors consistent with the findings of Experiment 3. We observed substantial levels of vertical errors in collinear, with half of the errors at levels significantly greater chance (targets t0, t2, t5, t6). Interestingly, of the 7 target items in *noncollinear* that had collinear vertical opportunities, all such opportunities were taken at least once, and three targets (t3, t4, and t6). We also saw collinear horizontal errors at levels significantly greater than chance in both collinear (t4, t5, and t6) and to a lesser degree in noncollinear (t6). Furthermore, the noncollinear condition showed some out-of-set (D) errors going to the same item in a set diagonal to the target set. One of these errors was significantly greater than chance (t0, $p$ = .019), and two others neared significance.

### 6.2.2 Cognitive load: Results of NASA/TLX

NASA/TLX scores are shown in Table 5. Participants were asked to report cognitive load levels for the brainstorming experiment only and not the recall test. The weighted total of the difference was not significant ($p$ = .778). Although we did not expect a difference, the Effort scale showed that participants felt that they exerted more effort while brainstorming in the collinear condition than noncollinear (51.6 vs. 43.3, $p$ = .044).

## 6.3 Discussion

One might voice concerns about the low mean number correct. However, onsidering this was a fully incidental, short-duration task with a secondary effortful task, low numbers like this should not be too surprising. By the same token, we must consider the fact that the noncollinear condition showed mean correct scores nearly twice as high with the same brief exposure, and that the correct location of 2 or more items was met by a large majority of participants in

Table 5. Experiment 4 Relocation Errors

| | | Chance / 35 | | | | | Test by type of error | | | | | Relocation errors | | | | |
|---|---|---|---|---|---|---|---|---|---|---|---|---|---|---|---|---|
| | | **WS** | **CV** | **CH** | **D** | **O** | **WS** | **CV** | **CH** | **D** | **O** | **WS** | **CV** | **CH** | **D** | **O** |
| **COLLINEAR** | **t0** | 2 | 2 | 4 | 2 | 6 | —— | .000*** | —— | .000*** | .993 | 0 | 12 | 0 | 12 | 3 |
| | **t1** | 2 | 1 | 4 | 1 | 5 | .138 | .751 | .177 | .751 | .139 | 5 | 1 | 8 | 1 | 10 |
| | **t2** | 2 | 3 | 4 | 2 | 8 | .294 | .007** | .654 | .000*** | .998 | 4 | 10 | 5 | 10 | 4 |
| | **t3** | 2 | 2 | 4 | 2 | 8 | .055 | .138 | .177 | .138 | .977 | 6 | 5 | 8 | 5 | 6 |
| | **t4** | 2 | 2 | 4 | 2 | 8 | .000*** | .522 | .042* | .522 | .977 | 10 | 3 | 10 | 3 | 6 |
| | **t5** | 2 | 1 | 4 | 1 | 5 | .055 | .002** | .018* | .002** | .378 | 6 | 6 | 11 | 6 | 8 |
| | **t6** | 2 | 1 | 3 | 1 | 3 | .055 | .048* | .007** | .048* | .791 | 6 | 4 | 10 | 4 | 3 |
| | **t7** | 2 | 1 | 3 | 1 | 3 | .522 | .751 | .598 | .751 | .394 | 3 | 1 | 4 | 1 | 5 |
| | | 16 | 13 | 30 | 12 | 46 | | | | | | 40 | 42 | 56 | 42 | 45 |
| **NONCOLLINEAR** | **t0** | 2 | 1 | 2 | 2 | 7 | —— | .751 | .768 | .019* | .998 | 0 | 1 | 2 | 7 | 3 |
| | **t1** | 2 | 1 | 2 | 1 | 5 | —— | .751 | .522 | .751 | .700 | 0 | 1 | 3 | 1 | 6 |
| | **t2** | 2 | 1 | 2 | 2 | 8 | .768 | .400 | .768 | .941 | .944 | 2 | 2 | 2 | 1 | 7 |
| | **t3** | 2 | 2 | 0 | 2 | 6 | .055 | .019* | —— | .055 | .302 | 6 | 7 | 0 | 6 | 10 |
| | **t4** | 2 | 2 | 0 | 2 | 7 | .294 | .006** | —— | .055 | .148 | 4 | 8 | 0 | 6 | 13 |
| | **t5** | 2 | 1 | 3 | 2 | 4 | .138 | .400 | .986 | .294 | .979 | 5 | 2 | 1 | 4 | 2 |
| | **t6** | 2 | 2 | 2 | 3 | 4 | .000*** | .019* | .006** | .986 | .091 | 10 | 7 | 8 | 1 | 9 |
| | **t7** | 2 | 1 | 1 | 1 | 2 | .001** | .751 | .157 | .400 | .941 | 9 | 1 | 3 | 2 | 1 |
| | | 16 | 11 | 12 | 15 | 43 | | | | | | 36 | 29 | 19 | 28 | 51 |

WS = Within set, CV = Collinear vertical, CH = Collinear horizontal,
D = Different set, O = Other; **p* < .05, ***p* < .01, ****p* < .001.

noncollinear (compared to fewer than one-third of collinear participants). This implies a memory advantage for noncollinear layouts that is consistent with the results of Experiment 3 and the first two experiments. In addition, the specific nature of the location errors offers some further demonstration of a collinearity confoundment effect resulting from aligned distractors or aligned structures in general. There are higher-order, even more realistic activities on digital devices that can probe these phenomena better than this one; however, we believe Experiment 4 carried the chain of custody from the initial memory traces seen in contextual cueing, up to an experiment with adequate environmental naturalism.

Table 6. NASA/TLX Results (0-100 scale, $N$ = 96)

| | Collinear | | Noncollinear | | Overall | | |
|---|---|---|---|---|---|---|---|
| | ***M*** | ***SD*** | ***M*** | ***SD*** | ***M*** | ***SD*** | ***p*** |
| **Weighted total** | 41.0 | 22.1 | 42.2 | 17.7 | 41.6 | 19.9 | .778 |
| **Mental** | 42.3 | 25.1 | 46.4 | 24.8 | 44.3 | 24.9 | .427 |
| **Physical§** | 22.5 | 20.5 | 18.9 | 14.0 | 20.7 | 17.6 | .312 |
| **Temporal** | 45.4 | 26.9 | 41.6 | 21.9 | 43.5 | 24.5 | .444 |
| **Performance** | 38.8 | 26.2 | 41.6 | 23.6 | 40.2 | 24.9 | .582 |
| **Effort** | 51.6 | 18.3 | 43.3 | 21.0 | 47.5 | 20.0 | .044* |
| **Frustration** | 33.4 | 26.9 | 34.8 | 25.8 | 34.1 | 26.2 | .802 |

§Physical scale failed Levene heterogeneity test ($p$ = .041, skewness = 1.5, kurtosis = 3.1). *$p$ < .05

# 7 General discussion

The CC experiments Given that CC's effect over time is to help guide attention and facilitate search (Chun & Jiang, 1998), it should follow that the collinear arrangements depressed the CC effect and must lack some affordance that the noncollinear arrangements enjoy. The third two experiments the fourth experiment

Any putative RT advantage for visual search on collinear arrangements was nullified by the small amount of practice with the more memorable noncollinear arrangements. Reaching

We believe that the ecological chain of custody was followed with reasonable care in this series of experiments. The experiments appear to demonstrate that the contextual difference was behind the disadvantage for collinear arrays in the designs. Some results suggest the likelihood that this consistency in object arrays may in fact *give rise to* item confoundment and additional effort. This may not seem substantial until we place it in the setting of a typical user. Let us take the contextual cueing effect numbers conservatively as one (admittedly imprecise) indication, and imagine a typical user needing to dedicate on the order of 70 to 100 ms to selective feature search within a collinear array – rather than some other, more automatic action that may ultimately take the same amount of time but will involve less cognitive effort. This fraction of a second may seem quibbling. And yet we engage in such microsearches in tightly arrayed sets of controls, menus items, and other screen objects almost constantly while using our devices. Furthermore, our busiest experimental display had only 35 distractors; it is not atypical of consumer and work screens today to have hundreds of distractors on one or more consecutive panels of material. It can be argued that on digital devices, whenever there is no sustained attention on a task such as viewing or composing, the main other possible task is combing through distractors for resources. An equal concern is that perusing and composing content are the precise activities where we most need to preserve our cognitive resources to be absorbed in work, to limit extraneous cognitive load through non–task-relevant activities (Sweller, Van Merrienboer, Paas; Lavie, 1995).

Prodigious memory for imagistic scenes Shepard 1967, Standing 1973, Brady Konkle Alvarez (see LXQ wb sl. 60)

If these numbers from the laboratory generalize, then work activity on current digital devices involves effortful search due exclusively to the design standard of collinear arrangement of screen objects, while noncollinear designs would present opportunities for reach affordances during the same work that may nearly entirely avoid this expenditure of effort. The results here show search tasks among one to three dozen or so items on screen. Activities regularly involve extended work going through much more material, whether on a single screen or requiring panning, scrolling, and zooming. Any location uncertainty will trigger an effortful feature search throughout the task, as we see in the photostream scrolling example and as should be apparent from everyday anecdote. The effort continues until either the target is found or distractors are exhausted and search fails. In contrast, given the reportedly prodigious imagistic memory available in a reach-style environment (e.g., zooming on a map of a familiar area for a familiar site), a spatial reach activity for an object whose location is well known would involve very little focused attention, perhaps only that applicable to Fitts' law (for reviews, see Chakraborty & Yadav, 2022; R. Jiang & Gu, 2020). For compound activities in the real world of several steps and/or in large selection spaces, the difference in effort may often be substantial, amounting to a much larger part of the hour for activities involving search. The beginnings of a computation for cognitive load would rely for factors on the number of targets; number of distractors globally; number of screens to traverse; size of targets relative to screen size; and (if the current results generalize) number, proximity, and density of local collinear distractors.

Our studies attempt to maintain an experimental chain of custody between clear demonstrations of CC at more environmentally authentic levels. Consistent with CC studies, subjects correctly reported no greater recognition familiarity than at chance levels with either displays or object locations in the CC "old" condition. However, in the higher-environment experiment of in Experiment 2A, both aligned and eccentric conditions registered accurate location recall scores, and these memories survived well over a week's time. Presumably, this was due to the more repetitive training and far stronger familiarity with the single display than subjects had with the 372 different CC displays. The chief difference along the experimental chain of custody that would explain the failure in CC is some fact of interference by the CC "new" displays. The 372 displays are temporospatially superimposed over one another, causing an interference of occlusion. Instance theories of attention and memory, vvv hold that multiple features (), multiple instances or even all instances are encoded in memory. Logan – whose CTVA theory of automaticity and memory retrieval is used to explain CC – held that "Automaticity *is* memory retrieval: Performance is automatic when it is based on single-step direct-access retrieval of past solutions from memory" (Logan, 1988, p. 493). Although it is evident that in CC "old" instances are reinforced, the "new" instances are also at least partially encoded: the task's time constraints are complicated further by the spatial constraints, which place all instances superimposed upon one another. In our CC experiments, the perfectly consistent occlusion in the "new" collinear arrangements contributed to their failure to match the retrieval efficiencies of the noncollinear arrangements. Following that, in Experiment 2A, the number of correctly located butterflies was consistently higher for the noncollinear arrangement across the week's four recall tests. Although on Day + 0, with the exercise completed just moments before, the difference was only marginal, the advantage for noncollinear increased substantially with the passage of time, and vertical and horizontal collinear confoundment errors also increased substantially. In other visual experiments, items have been randomly jittered to prevent anticipation of location (Richardson, Spivey, Barsalou, & McRae, 2003, p. 771) and collinearity (van Zoest, Giesbrecht, Enns, & Kingstone, 2006, p. 536). We deliberately did not jitter items; yet, counter to what some experimenters might expect, this in fact was a disadvantage for the collinear condition. The sole manipulation was collinearity, and so one possible explanation from an instance theory account is that encoded instances with some spatial variation, despite rapid presentation and similar cognitive load, are less occlusive of one another than in matter that superimposes.

It is a concern that this the type of display occlusion interference described above is now a constant in HCI, for the very reason that screens must accommodate large numbers of objects and that programmers have decided that perfect alignment is the most efficient way to achieve this. Today's app menus is an everyday case demonstration of multiple facets of occlusion. Given the identical background design and background color on three or more adjacent screenfuls of icons, when side-swiping to locate an app, users must often resort to a feature search simply to clarify which of several screens they are on, before initiating a *second,* subsequent feature search for the desired app (Figure). This all perhaps helps to explain why we can remember only approximate locations of objects in these rectilinear arrays, even when using only a small number of screens, and also why the confoundments are predominantly to the next vertical and horizontal neighbors.

The Conci and Kimchi labs' results assert that percepts of closed objects may form perceptually closed, hierarchically superior objects of their own and, in doing so, may interfere with spatial contextual learning. Chun and Jiang (1998, p. 40) also assumed that instance-based representations do not preserve information irrelevant to a given task, and yet their own results demonstrated that location is relevant. The mind will compute collinear contours into larger percepts that can influence contextual learning. T/L structures, Julesz shapes, Gabor patches, and other special stimuli used in psychophysics experiments create strange hybrid objects notorious for messing with figure and ground interpretation and leading to metastability (Wilson & Keil, 1999, p. lxii). The Kimchi and Conci experiments forced this point by creating the illusion of new objects and causing contextual confusion. However, the borders around our objects (and therefore screen icons) partially or completely pre-empted this possibility, even in the case of the (enclosed) T/L objects of Experiment 1. However, an exception appeared when groups of collinear objects created linear forces of their own to cause confusion (cf. Jingling & Tseng, 2013; Tseng & Jingling, 2015).[3]

Brockmole (2006) found that contextual cueing worked when global locations remained constant and local features changed, but not the other way around, with local features staying constant and global location changing. And yet the latter describes the current state of the art of interface displays.

[3] This suggests concerns about the interpretation of any cognitive visual experiment whose local configurations are composed of open contours but where configural closure is not a phenomenon of interest. This clearly includes any experiment that relies on letters as stimuli, such as Navon (1977).

### 7.1 And, related: Is space a feature?

In the world of design, space is treated with greater respect than the objects themselves. Space may or may not be a feature in its own right. [Objects may be placed in space to constitute hierarchical groups (Navon, 1977).] Collinearity is occasionally deemed a quasi-feature shared and creating hierarchical unity among related objects (Pylyshyn, 2007, pp. 1-17). Noncollinearity may thus be considered the absence of the feature of collinearity. But as to irregular gaps that oppose collinearity, if the gaps themselves have differential qualities, then the distinction of space would seem to lie existentially outside of qualities attached to the targets and distractors, and the story of the shape of the void makes these voids, at minimum, carry featuristic properties, if not objects with features in their own right. Regardless, treating interstitial gaps as a global factor for search purposes means that collinearly arrayed display matter with great internal feature variation still may be surrounded by gaps that all share visual similarity (). This happens to be the state of the art in screen design today, where brand icon designers are constrained to competing amongst one another within their assigned borders, and photographs in photo-streams and menu items do the same, while the gaps between the objects are defined and fixed by operating-system programmers. These gaps, uniform as they typically are on our devices, may or may not be features. Calling space a feature may complicate things a great deal. What, for example, does space bind to? It would also be problematic to treat spatial gaps as distractors. One important concern here would be reconciling this assertion with the general principle that increasing distractor differences hinders target search and increasing distractor similarity fosters target search (Chun & Wolfe, 2005; Duncan & Humphreys, 1989). The opposite is true in the case here. And yet these experiments suggest the notion of *nontarget space,* space in the role of distractor. The uniform gaps in collinear arrangements seem to have served this function (Figure), subject to the rules of non-target stimuli and conjunctions. It should be noted that in our experiments the CC effect was still apparent despite our having put uniform borders around objects, ostensibly making them effectively all identical in shape. Without insisting that space is a feature, this fact nevertheless suggests that the featuristic value of a uniform space may often be stronger than the *bona fide* feature value of the inter-object uniformity of the border enclosure. This featuristic value afforded visual richness serving memory in our noncollinear conditions just as unique targets might be expected to do, but this richness was lost in the collinear conditions. Since search speed is dependent on target-distractor differences Another way to view this is that in noncollinear, interstitial space target-distractor Experimentation in the area of conjunction feature search and contextual cueing would benefit from a closer study of the shape of the space created between objects, as well as the relationship between the salient contours of closed objects and the extended lines implied by these contours (Kimchi et al., 2016). In any case, we have found some explanatory value in interpreting space as a limited kind of feature of the objects, and the fact that memory for Experiment 3 has demonstrated that

There is also evidence in CC (Brockmole) and neural research (Li Jingling & Tseng) that perceptual regularities impede spatial learning. If this

## 7.2 The Bonsiepe collinearity doctrine

Experiment 3 showed, that search speed in collinear layouts was no faster than noncollinear, initial error rates were higher, and recall accuracy suffered significantly greater impairment with the passage of time than for noncollinear layouts. Furthermore, there was some evidence that the the collinearity itself triggered location recall confoundment with immediately vertical and horizontal items. Folk theoretical claims in favor of the efficiency of rectilinearity seem have early inspiration from the Shannon-Weaver (1949) entropy law in information theory, prescribing compact, low-entropy information structures for efficient information transfer of all sorts. It must be conceded that a presumption in favor of low-entropy visual displayed matter did make screens appear tidier. For the pre-GUI era, with limitations of 80-column ASCII and ANSI text displays, the belief that it could improve usability might have been understandable. It was particularly welcome by computer scientists, who were primarily neat, rational thinkers, not designers or psychologists accustomed to studying human needs. The entropy law also had a contemporary companion assumption: the idea that humans are inherently information processors (Card, 1982; Card, Moran, & Newell, 1983) fit well with propositions that efficiency of interaction with computers could be reduced to a simple problem of bit density.

The moment that collinearity became gospel for scientific design can be dated fairly accurately. Naturally, it had already been accepted in the design psychology community as conforming with the rules of Gestalt Prägnanz. But claims of its advantage for screen layout appears to stem from an influential paper by Gui Bonsiepe (1968) relating to print design. Bonsiepe had begun following information entropy theory several years before PARC programmers took it up. He devised a formula based on the Shannon-Weaver (1949) information entropy law:

$$C = -N \sum_{i=1}^{n} p_i \log_2 p_i$$

where *C* is the measure of information complexity, upper-case *N* the number of screen objects, lower-case *n* the number of similar object groups, and $p_i$ the probability of selecting an object from group *i* (Bonsiepe, 1968; Tullis, 1981). Bonsiepe used the formula to compute the efficiency advantages of a catalog page-grid redesign (Figure 13). The now-famous example demonstrated that the new version was 55% “simpler” than the old – at least in terms of information theory. Tullis paid particular attention to addressing Bonsiepe’s formula for computer display organization and text-block alignment for improved usability (Tullis, 1981, 1984).

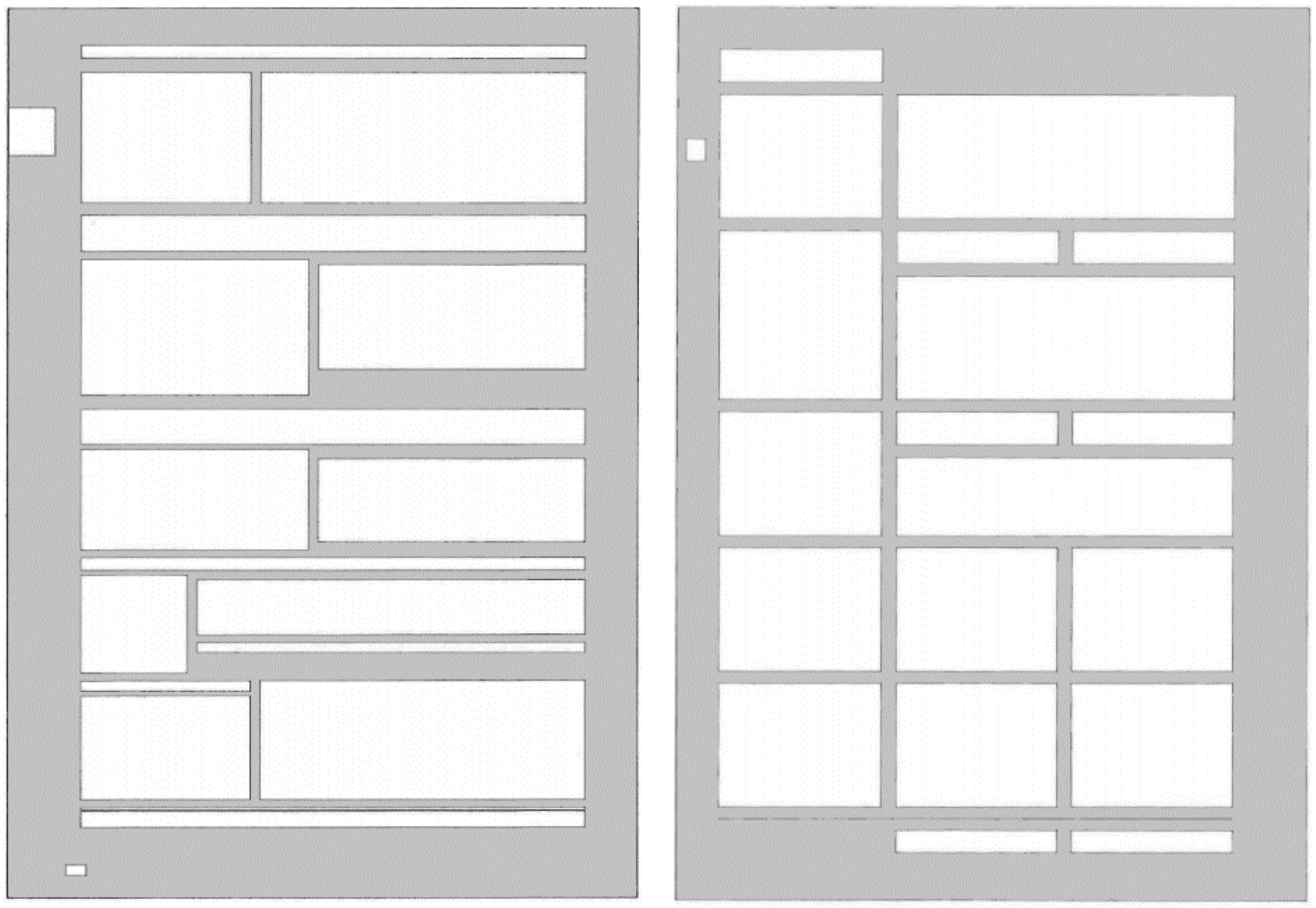

Figure 13. The results of Gui Bonsiepe’s (1968, pp. 210–211) application of the Shannon-Weaver entropy law to design. Left, a typical catalog page. Right, the computed lower-entropy rendition. (Reprinted with permission of *Visible Language*, Mike Zender, University of Cincinnati.)

In screen layout, a presumption of the advantages of collinear arrangements for GUIs can also be dated fairly accurately to 1982 or 1983. This is regarded as the birthdate of HCI, coinciding with the debut of the Apple Macintosh, the creation of the Conference on Human Factors in Computing Systems (later ACM SIGCHI), and the publication of the book *The Psychology of Human-Computer Interaction* (Card et al., 1983; MacKenzie, 2013). Designers began taking a more wholesale interest in screen design only around that moment. Bonsiepe’s idea was attractive to computer programmers. They accepted without critique a transmission of the information entropy rule from print design, through Tullis’ ANSI text demonstrations, in turn to aspects of GUI layouts (Comber & Maltby, 1996; Parush, Nadir, & Shtub, 1998). In the new visual vocabulary of the GUI, screen icons were generally squarified targets, to be selected by pointing, and representing both noun- and verb-type screen objects of any sort. They were made to rectangular standards so as to allow for efficient coding in the operating system and rectilinear arrangement measured by the extents of the display.

The problem with we are branding the Bonsiepe doctrine is that, even assuming that it is mathematically and philosophically more or less sound, it limits the target-acquisition problem to one of pure visual search speed, without any accounting for memory and other endogenous processes. Hence much relies on internal features of the object such as color and other salient features. While goal-directed processes may improve speeds even on internal features, the general emphasis on such designs is on affording bottom-up microsearches. The objects themselves are left to be auto-arranged in rectilinear patterns based on the principle of low entropy, presumably for rapid scanning. Yet this ignores how higher-entropy arrangements of the objects may actually serve memory better.

This strategy causes display design to rely on the users' goal-directed object search. Many objects in display layouts have highly similar features such as color, outline shape, and graphical similarities. Consequently, users rely on internal feature saliency. To avoid this cognitively costly strategy, users will often solicit help from excitatory/inhibitory filters (Rensink & Enns, 1995, p. 102). One example – likely enacted on phones tens of millions of times a day – is to find the Facebook app icon on our phones. We do this not by remembering that it is within easy reach in, say, row 2 of column 3 of screen 2 of our app menu, but by thinking of a bold lowercase "f" knocking out of a large blue background among identically shaped, collinearly arranged rectangles. These are one or two separate excitations; or, alternatively, perhaps one or more inhibitions – although training will enhance performance, mostly supported by a series of visual search schemas.[4] This may seem to be relatively fast and non-problematic, but it is also an effortful dedication of sustained focused attention, and moreso when under stress. In another everyday example, imagine spending 15 to 30 seconds scrolling through our photostream to show a friend an interesting photograph. Immediately after closing the activity, a second friend asks to see it, calling for nearly identical labor, again moreso when under stress. Kahneman wrote that "…distraction is resisted at a cost: motor tension and autonomic manifestations of arousal are higher than normal" (Kahneman, 1973, p. 113). Here we must remember that even the stress may cause distraction, but in attention-based feature search, distractors are also a distraction.

Relying only on efficient visual search fails to account for the problem that the lower-entropy entities will necessarily form uniform higher-level entities that leave little for memory to latch onto. Reaching for a thing, on the other hand, can be said to be virtually absent of distraction. Reaching is automatic, proprioceptive (Pylyshyn, 2007, pp. 3-47 ff.), and not effortful. Recent findings in both experimental and neural areas are arguing that the ecology of the "reachspace" is entirely different from that of individual objects or full scenes (Josephs, Hebart, & Konkle, 2023; Josephs & Konkle, 2019, 2020).

### 7.3 The use of space in layout

The collinear and clustered arrangement of the collinear conditions in the experimental designs was intended to approximate the state of the art in interface design. However, this was in no way meant to suggest that the noncollinear conditions are specifically prescribed as quality design solutions. They were arranged in this way only as an expedient to experimental consistency among the conditions. Rather, designs emphasizing substantial inter-object and inter-group spatial variation as a design factor, should probably redound to greater reach opportunities, to reduce cognitive effort and time. It is true that a first step in typographic and web design is always to establish a grid. Primarily for text reading, this may be beneficial, though specifics are also frequently debated among graphic designers. Grids in menus, controls, and object spaces may sometimes be beneficial. But locating frequently accessed controls and work objects arrayed on a display, or any content in large arrays, might be better served using free arrangements.

[4] Users of WeChat or WhatsApp will filter for different logotypes on a *green* ground, but in any event the inhibitory/excitatory formula and cognitive demand for the selective input activity are essentially the same.

[[Kirby16 while ability to detect may lead to proclivity to organize, it does not follow that neatness organized information is always easier to access; however, this may be the logic that designers use]]

If the thesis presented here were to prove to have some credibility, the notion of "pop out effect" may need to consider inter-object rectilinear structures as having one kind of preattentive salience (i.e., from emergent collinear proto-contours) that may inadvertently trigger attentional capture and interfere with search and recall, and nonrectilinear spatial regions as having a quite different kind of preattentive salience (i.e., from their richness of spatial variety) that may serve search and recall.

Space is increasingly at a premium on digital displays. How to interact with growing amounts of content within a small space is of great concern to designers and users. An important question in display design is what to do when the screen becomes too crowded. The two main available options are to add another screen or to shrink content size, selectively if possible. Adding another screen is often disorienting, and doing so also preempts opportunities for all content for a given task to occupy a single work-plane. Shrinking content and other strategies, such as lens effects, have been proposed on many occasions (for overviews, see Healey & Enns, 2011; Keim, 1997, 2002). These are more often proposed for data visualization, less often for general display screen use. Few alternative proposals so far appear to be of practical value. In any event, most modern screen layouts do not attempt to preserve location state even on single screens of content, but instead automatically move objects around at will with no consideration for location memory. This includes most of the alternative proposals mentioned. Using a neural-network model to predict further evidence from human subjects, Brady and Chun (2007) found that stable local context is necessary for contextual cueing to take effect. Using N1pc evidence, Zinchenko and colleagues (2020) recently showed that moving a target to the opposite hemifield abolishes contextual cueing. An example of this kind of disturbance in real environments is when someone rearranges a room in one's home and it takes time to adjust to the change. In interface design, despite injunctions against it (Shneiderman & Plaisant, 2005), this class of displacement is quite routine, done systematically when redesigning applications (Figure 14) or even automatically by the operating system on a moment-to-moment basis.

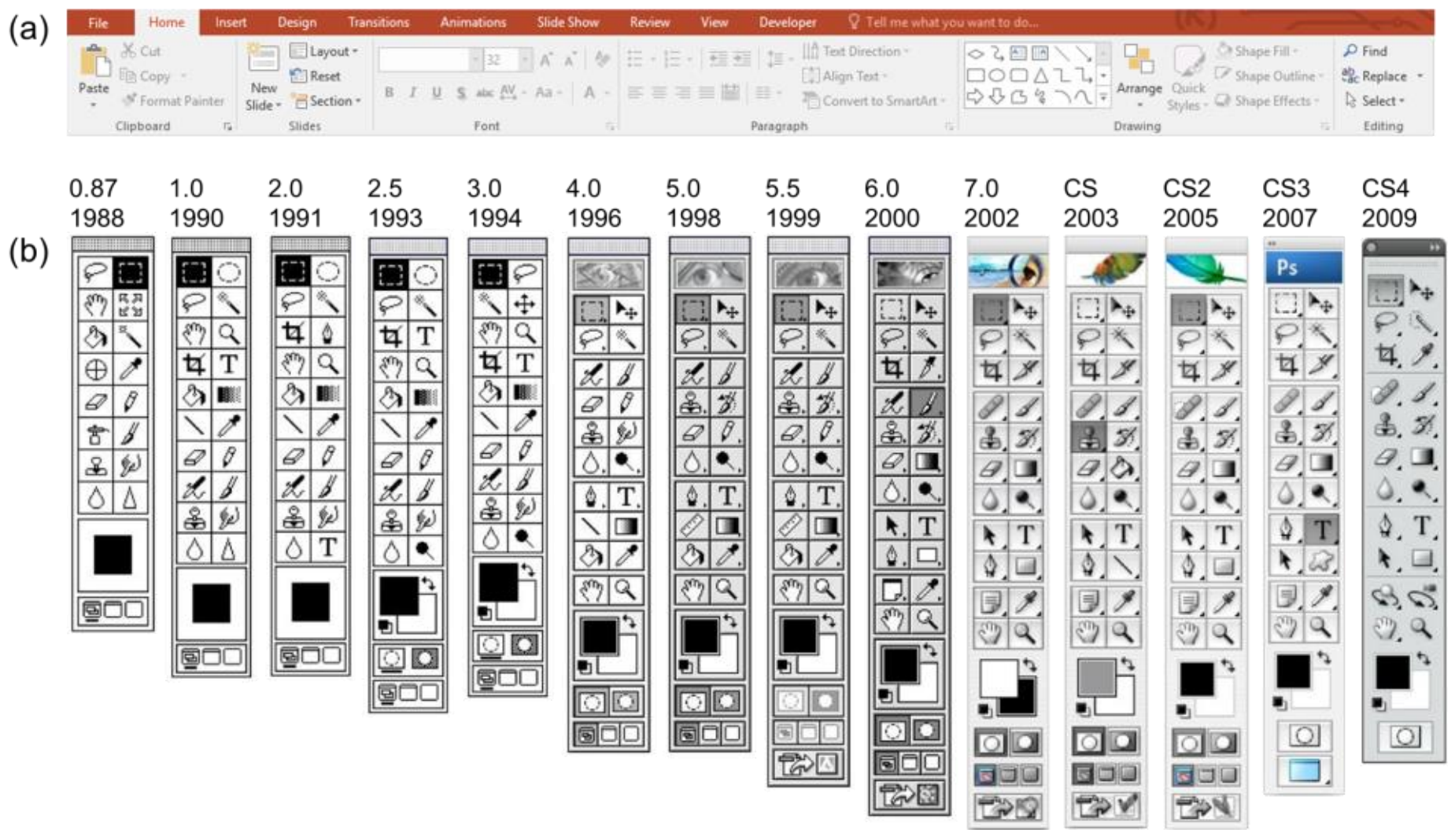

Figure 14. Examples of the alignment standard in menu interface design of popular applications show opportunities for vertical and horizontal object confoundment. (a) Microsoft "Fluid" horizontal ribbon menu. (b) Adobe Photoshop vertical toolbar menu changed 13 times over a 21-year period, with each version adding features and swapping feature locations.

Competitive icon coloring and other contrasting internal feature elements within the icons makes the start of a schematic race between spatial reach and a visual search; however, our demonstration here suggests that reach should ultimately be more cognitively efficient, if given its proper affordances of space by not imprisoning objects within a grid. This is to say that new display designs that maintain previously accessed items at or near nonrectilinear locations, and without strongly changing local context, should lead to less confusion and reduced visual search.

### 7.3.1 A defense of 2D reachspaces over 3D spatial screen work

We suspect that 2D interfaces are fully capable of fulfilling virtually any typical human-computer interaction need, and we believe that design technique has not come anywhere near to exhausting the opportunities of the 2D graphical interface. Yet some concerns exist regarding the practical future of 2D interaction research, and we believe it would timely to address these concerns. Research over more than 30 years has gone into studying three-dimensional and other novel interface designs, nominally as an aid for data visualization but also for general interaction (Healey & Enns, 2011; for many examples, see Keim, 1997). Although there may not be a consensus, there appears to be a long-held view among many HCI practitioners that supplanting 2D interfaces with 3D interfaces is a question only of time, and that this inevitable fact may at length render research and practice on planimetric interfaces irrelevant. Let us set aside the point that no coherent reason has so far emerged for 3D interfaces to become the norm over 2D. Aside from this fact, a concern should be stated at the outset regarding different user populations. Importantly, gender and age may be in play (Gutwin, Cockburn, Scarr, Malacria, & Olson, 2014; Hubona, 2004; Postma et al., 2004; Scarr, Cockburn, & Gutwin, 2013; Silverman, Choi, & Peters, 2007; Walkowiak, Foulsham, & Eardley, 2015). Evidence is mixed regarding differences between men and women in search and memory tasks; however, it is increasingly conceded that men tend to excel at mental rotation, sense of direction, and consequently survey-style navigation (Postma et al., 2004; Walkowiak et al., 2015). On the other hand, evidence is also accumulating that shows smaller differences between men and women in route knowledge than previously believed. As to older users, there is evidence that while attentional acuity and explicit memory degrade with advancing age, implicit memory for spatial relations is not affected (Hasher & Zacks, 1979; Postma et al., 2008, p. 1343). The ability to recall the precise position of objects was not greatly impaired for 80-year-olds when compared to 20-year-olds in one study (Pertzov, Heider, Liang, & Husain, 2015). This evidence would appear to favor new interaction designs that encourage implicit memory strategies rather than attentionally demanding search strategies.

That said, adding a dimension for both orientation and manipulation may do more harm than good, and so far we have seen are no adequate justification for it in HCI. Studies for new interface designs using a representative segment of the anticipated user population are scant. Studies on spatial interfaces frequently draw most of their subjects from expert technical populations, groups of young, male computer science students (Cockburn, Kristensson, Alexander, & Zhai, 2009; Cockburn & McKenzie, 2004), and technologists (Kandogan, Kim, Moran, & Pedemonte, 2011). In addition, although HCI studies do at times make efforts to cite evidence from cognitive psychology, the studies themselves tend to resort to pitting novel concepts against existing straw-man designs that anecdotally are already known to perform quite poorly. Perhaps anything would be an improvement on that status quo, but that does not necessarily mean that the proposed designs were carefully considered for populations other than experts. In any event, studies from HCI regarding object-location memory are mixed. An early study found an advantage for location memory when using 2½-D presentations (Tavanti & Lind, 2001). Much research in this area comes from Cockburn and colleagues. Two of their related studies, partly as a response to Tavanti and Lind, found no advantage for spatial memory for 3D designs over 2D designs (Cockburn, 2004; Cockburn & McKenzie, 2002). A third study by the same researchers found location memory better for a 2D model than a 3D physical model (Cockburn & McKenzie, 2004). These early studies did not test in immersive or stereoscopic environments, although the Cockburn group cleverly used physical models in anticipation of the same. A separate study found that larger 3D displays did bring improvements to spatial memory (Tan, Gergle, Scupelli, & Pausch, 2004). After this period, HCI research on 3D interaction appears to have begun to deviate from the earlier comparative studies with 2D regarding GUI environments and ease of use (including location recall), and to move towards 3D visualization, gaming, and wayfinding applications. References to relevant cognitive research also dwindle.

Regarding direct manipulation interfaces, in a widely used model, Hutchins and colleagues (1985, p. 318 ff.) used the gulfs of execution and evaluation to describe the distance among goal, action, and system feedback in human-computer interaction. They further distinguished semantic and articulatory communication distances in HCI (p. 321 ff.) for traversing these gulfs. *Semantic* distance refers to how closely a user's intent tallies with the various possible ways to indicate this intent within a given user environment. *Articulatory* distance refers to how closely the specific linguistic tokens (or action metaphors) of the user's language tally with the machine's language (or response metaphors). The idea was to persuade programmers of the benefits of providing the most direct means of communication in both providing (semantic) concepts and (articulatory) metaphors for the concepts to allow users to most directly manipulate the available tools to accomplish goals.

Let us consider 2D versus 3D graphical user interfaces within this semantic-articulatory frame for interaction distance. It may be acknowledged that humans live in 3D. However, let us further assume that Josephs and Konkle's evidence for the existence of a cognitively distinct, more automatic "reach" type of human schematic interaction may be valid, and also consider the varying needs of different populations of users as listed above. All other things being equal, we believe that it is likely that most object tasks on digital interfaces would be considered more universally proximal when treated as 2D reach work, by measure of both semantic and articulatory distance in direct manipulation interfaces, without the added work of rotation or other spatial gymnastics. This may even be the reason for the longevity of the 2D interface and its pointing devices, beyond their more modest technical requirements (which should also be taken into account). It should further follow that aside from perhaps its employment as an expedient in rare cases to help navigate, manipulate, and interpret some complex data or resolve certain specialized problems, 3D interfaces will usually exceed the use general case for everyday object interaction for most users except perhaps those with the physical abilities, semantic skills, and practical need for such added dimensionality, for example gamers and high-level data analysts.

Goujon and colleagues' detailed review (2015) remarked that CC "only tolerates very limited displacements of the items and remains viewpoint-dependent, position-specific, and absolute to the configuration" (Goujon et al., 2015, p. 528). Therefore, they conclude, "both the local context of the target and its place within the global context are crucial for CC" (Goujon et al., 2015, p. 528). Taking this in context with how we both manipulate and forget objects in both virtual and real space, it appears that any *excessive* displacement will cause interference. Therefore, it should follow that *any* interface (spatial, tactile, auditory, etc.) interested in facilitating memory-based retrieval should make special efforts to minimize distortion of the interstices and perceptual objects contained in the interfaces.

Given all of these points, if we continue to use computer displays primarily to access everyday objects and perform everyday human activities in simple object:verb icon selection relationships (e.g., “these six blurry photos:delete”) using hands and related pointing metaphors, then the vast majority of our graphical interface interactions would be appropriate in the form of arm’s-length reach, two- or (occasionally) two-and-a-half-dimensional work, and should rarely if ever need to resort to three- and four-dimensional interaction modalities on most devices. How this relates to general human wellbeing should be implicit from the true meaning of ease of use. That we trace this error in theory back to Bonsiepe of the Ulm school is not problematic, since they understood the weighty role of the designer and it is likely they would have altered this part of their thinking if they had been aware of the larger concerns.

## 7.4 Limitations and further work

It is not fair to assert from these results whether collinearity induces errors or that relaxing the spatial constraints improves memory. Identifying the nature of this differentiation is essential.

We believe collinearity in digital device display designs is one culprit driving search inefficiency and increased cognitive load due both to memory encoding and retrieval. The later experiments lacked sufficient opportunities to clarify whether incidental collinear errors have influence in noncollinear arrangements. There were hints of this influence, but more is needed to tie these practical environmental experiments to the CC experiments that demonstrate some interference from emergent collinear structures (Conci & von Mühlenen, 2011; Jingling & Tseng, 2013; Kimchi et al., 2016; Tseng & Jingling, 2015; Zinchenko et al., 2020).

More experiments in the contextual cueing genre testing fewer “old” displays with more naturalistic occlusion of fewer “new” displays would create a more complete chain of custody to demonstrate the likelihood that occlusion is the cause of the failure of recognition memory in CC experiments. This similarity occlusion, we believe, strongly influences both within-display and inter-display memories. Related to this is the extent of distortability of local and global space, a particularly important phenomenon to measure. Work may also attempt to further identify the nature of and quantify the extent of the impact of rectilinearity on attention and memory in applied interaction.

# 8 Conclusion

The methodologies used for these experiments are not an approach to or from design, nor are our experiments prescriptions for good design. Rather, as a group they serve as a useful demonstration of how excessive uniformity in design can be a hindrance to users. The findings happen to accord quite well with use cases for 2D reach environment designs. Hence we can argue for the development of GUIs that promote memorability of the reachspace by preserving local spatial state with configurally rich arrangements rather than rectilinear ones that offer no purchase for memory; and for GUIs that respect the global spatial arrangement of a virtual workspace. Within this frame, modest transforms of the global view can be tolerated as long as relative positions of objects are more or less maintained.

Earlier, we showed an influential historical prescription in HCI declaring that columnar structures in GUIs are beneficial for framing attention and for promoting scanning, and that uncluttered layouts “reduce competing stimuli for user attention” and endorse the use of “columnar structure” (Sutcliffe & Namoune, 2008, p. 18). We have offered initial evidence that rectilinear structures may inhibit good location memory, and in fact that that the rows and columns themselves appear to contribute to confoundments to objects immediately above, below, and to the sides.

A final point concerns practice. All modern operating system and web programming stacks implement essentially systematic rectilinear presentation of all content at the user-facing end. Not only is content poured in and maintained using grids, avoiding reliance on the grid is difficult and sometimes impossible. While more random-access display environments are implemented and exist, they are almost always built atop these rectilinear frameworks and must circumvent the default supports beneath them. This makes alternative proposals rare and difficult to manage, even aside from the basic fact of the predominance of the rectilinear doctrine.

It is hoped that this series of experiments contributes to the advancement of both theory and practice. It offers a series of demonstrations that the ability of the mind to latch onto unique visual arrangements is an aid to memory and subsequent easy relocation of objects, and that collinearity specifically appears to be implicated in hampering this capability. In practice when we design visual layouts, the linear aspect of groups of icon arrangements for menus, controls, and other objects apparently leads to confusion and consequent microsearches for lower-level features. This likely impacts object selection in a significant part of the use of display devices. It is hoped that it is not lost on the reader that this question of context failure leading to microsearches, should also apply to the rectilinearly constructed environment, including not only elevator buttons also the rows and columns of the apartments themselves. It is left to quantify how cognitive load is affected. We also hope that this series demonstrates one workable method for helping to bridge the historical gap between psychology and design.

# 9 Acknowledgments

We are grateful to Alan Baddeley, Nelson Cowan, Emilie Josephs, Timothy McNamara, Donald Norman, Andrew Ortony, Denis Pelli, Matthew Rossano, Jeremy Wolfe, and Xuelian Zang for general advice as well as comments on the manuscript. Thanks also to Manuel Charlemagne, Shane Johnson, and Peisen Huang for assistance during development. We also extend thanks to Yongqiang Huang, Xinyu Liu, and Huang Xueqi for diligent support in experiment management.

# 11 Appendix

DRAFT – DRAFT – DRAFT – DRAFT – DRAFT – DRAFT – DRAFT

| Experiment 1 | | | | | | | |
|---|---|---|---|---|---|---|---|
| | | RT | | | ER | | |
| Factor | df | *F* | *p* | $\eta^2_p$ | *F* | *p* | $\eta^2_p$ |
| LG | 1,60 | 6.85 | 0.01 | 0.102 | 10.3 | 0.002 | 0.146 |
| Context | 1,60 | 60.71 | <.001 | 0.503 | 24.99 | <.001 | 0.294 |
| Epoch | 5,300 | 116.78 | <.001 | 0.661 | 31.99 | <.001 | 0.348 |
| LG×C | 1,60 | 26.31 | <.001 | 0.305 | 11.94 | 0.001 | 0.165 |
| LG×E | 5,300 | 3.3 | 0.003 | 0.052 | 2.75 | 0.019 | 0.044 |
| C×E | 5,300 | 5.85 | <.001 | 0.089 | 2.07 | 0.068 | 0.033 |
| LG×C×E | 5,300 | 3.91 | 0.004 | 0.061 | 0.71 | 0.611 | 0.011 |

| Experiment 2 | | | | | | | |
|---|---|---|---|---|---|---|---|
| | | RT | | | ER | | |
| Factor | df | *F* | *p* | $\eta^2_p$ | *F* | *p* | $\eta^2_p$ |
| LG | 1,72 | 0.12 | 0.726 | 0.002 | <.01 | 0.955 | <.001 |
| Context | 1,72 | 33.94 | <.001 | 0.32 | 5.2 | 0.025 | 0.067 |
| Epoch | 5,360 | 72.23 | <.001 | 0.501 | 1.64 | 0.147 | 0.023 |
| LG × C | 1,72 | 4.45 | 0.038 | 0.058 | 0.07 | 0.794 | <.001 |
| LG × E | 5,360 | 0.35 | 0.881 | 0.005 | 1.28 | 0.273 | 0.017 |
| C × E | 5,360 | 4.54 | <.001 | 0.059 | 1.28 | 0.271 | 0.018 |
| LG × C × E | 5,360 | 1.29 | 0.269 | 0.018 | 0.64 | 0.669 | 0.008 |

| Experiment 1 RT ms (SD) | | | | | | |
|---|---|---|---|---|---|---|
| | Epoch 1 | Epoch 2 | Epoch 3 | Epoch 4 | Epoch 5 | Epoch 6 |
| Collinear new | 2229 (293) | 2058 (238) | 1856 (240) | 1849 (224) | 1836 (222) | 1730 (191) |
| Collinear old | 2184 (333) | 2066 (295) | 1880 (212) | 1802 (205) | 1781 (216) | 1702 (188) |
| Noncollinear new | 2307 (261) | 2155 (226) | 2141 (241) | 2079 (255) | 2029 (265) | 1905 (241) |
| Noncollinear old | 2299 (260) | 2099 (246) | 2030 (216) | 1940 (238) | 1846 (212) | 1722 (228) |

| Experiment 1 ER % (SD) | | | | | | |
|---|---|---|---|---|---|---|
| | Epoch 1 | Epoch 2 | Epoch 3 | Epoch 4 | Epoch 5 | Epoch 6 |
| Collinear new | 6.09 (4.50) | 4.44 (3.17) | 3.58 (3.36) | 2.82 (2.72) | 3.27 (3.22) | 2.87 (3.57) |
| Collinear old | 5.82 (4.62) | 4.57 (4.36) | 2.91 (2.19) | 2.59 (2.47) | 2.37 (2.92) | 2.91 (3.72) |
| Noncollinear new | 10.39 (6.09) | 6.63 (3.87) | 6.27 (3.32) | 5.47 (2.96) | 5.73 (4.48) | 4.97 (2.93) |
| Noncollinear old | 9.59 (6.41) | 6.27 (4.45) | 3.99 (3.26) | 3.49 (2.75) | 3.05 (3.13) | 2.78 (2.64) |

| Experiment 2 RT ms (SD) | | | | | | |
|---|---|---|---|---|---|---|
| | Epoch 1 | Epoch 2 | Epoch 3 | Epoch 4 | Epoch 5 | Epoch 6 |
| Collinear new | 1340 (417) | 1263 (340) | 1209 (283) | 1150 (280) | 1108 (267) | 1098 (247) |
| Collinear old | 1336 (420) | 1253 (362) | 1181 (293) | 1114 (270) | 1102 (266) | 1071 (268) |
| Noncollinear new | 1386 (358) | 1273 (282) | 1230 (249) | 1217 (256) | 1138 (241) | 1120 (221) |
| Noncollinear old | 1359 (351) | 1271 (285) | 1199 (237) | 1135 (237) | 1094 (218) | 1070 (218) |

| Experiment 2 ER % (SD) | | | | | | |
|---|---|---|---|---|---|---|
| | Epoch 1 | Epoch 2 | Epoch 3 | Epoch 4 | Epoch 5 | Epoch 6 |
| Collinear new | 1.12 (1.81) | 1.04 (1.38) | 1.17 (1.56) | 1.26 (2.02) | 1.40 (1.64) | 1.22 (1.65) |
| Collinear old | 1.17 (2.29) | 1.04 (1.49) | 0.72 (1.00) | 1.04 (1.54) | 0.77 (1.28) | 0.95 (1.65) |
| Noncollinear new | 2.03 (4.90) | 0.95 (1.15) | 1.26 (1.59) | 0.63 (1.14) | 0.94 (1.28) | 1.31 (1.25) |
| Noncollinear old | 1.80 (4.45) | 0.81 (1.22) | 0.50 (0.77) | 0.77 (1.22) | 0.72 (1.08) | 1.31 (1.76) |

| Experiment 3 RT & ER | | | |
|---|---|---|---|
| | Epoch 1 | Epoch 2 | Epoch 3 |
| RT collinear | 1633 (286) | 1053 (333) | 799 (250) |
| RT noncollinear | 1587 (273) | 1072 (309) | 868 (232) |
| ER collinear | 13.88 (7.60) | 5.43 (6.10) | 2.02 (2.93) |
| ER noncollienar | 9.39 (5.53) | 4.71 (6.29) | 3.07 (4.71) |

| | Counterbalance (which Exp.?) | | |
|---|---|---|---|
| Label | 1769 | 1221 | 905 |
| Label | 1517 | 997 | 815 |
| Label | 1522 | 949 | 775 |

Table 7. Experiment 1 Three-way ANOVA

| | RT | | | | | ER | | | | |
|---|---|---|---|---|---|---|---|---|---|---|
| **Effect** | ***F*** | ***df*** | ***p*** | **sig.** | **$\eta^2_p$** | ***F*** | ***df*** | ***p*** | **sig.** | **$\eta^2_p$** |
| **Layout** | 6.84 | 1, 60 | .001 | ** | .102 | 10.29 | 1, 60 | .002 | ** | .146 |
| **Context†** | 60.71 | 1, 60 | < .001 | *** | .503 | 24.99 | 1, 60 | < .001 | *** | .294 |
| **Epoch** | 116.78 | 1, 60 | < .001 | *** | .661 | 31.99 | 1, 60 | < .001 | *** | .348 |
| **L × C** | 26.3 | 5, 300 | < .001 | *** | .305 | 11.94 | 5, 300 | .001 | ** | .166 |
| **L × E** | 3.3 | 1, 60 | .006 | ** | .052 | 2.74 | 1, 60 | .020 | * | .044 |
| **C × E** | 5.85 | 5, 300 | < .001 | *** | .089 | 2.07 | 5, 300 | .061 | | .033 |
| **L × C × E** | 0.91 | 5, 300 | .001 | ** | .061 | 0.72 | 5, 300 | .611 | | .012 |

†"New" versus "old" display condition. *$p < .05$, ** $p < .01$, ***$p < .001$

Table X. Experiment 3 within-circle hit probabilities, binomial *p* values for collinear relocation errors, and error counts

| | | Chance / 35 | | | | Day + 0 Binomial *p* | | | | Relocation Errors | | | | | Day + 1 Binomial *p* | | | | Relocation Errors | | | | |
|---|---|---|---|---|---|---|---|---|---|---|---|---|---|---|---|---|---|---|---|---|---|---|---|
| | | WS | CV | CH | OT | WS | CV | CH | OT | WS | CV | CH | OT | OUT | WS | CV | CH | OT | WS | CV | CH | OT | OUT |
| COLLINEAR | t0 | 2 | 2 | 4 | 2 | .570 | —— | .787 | —— | 2 | 0 | 2 | 0 | 0 | .002** | .616 | .020* | .570 | 7 | 1 | 7 | 2 | 1 |
| | t1 | 2 | 1 | 4 | 3 | .570 | —— | .904 | —— | 2 | 0 | 2 | 0 | 2 | .010* | —— | .010** | —— | 6 | 0 | 9 | 0 | 5 |
| | t2 | 2 | 3 | 4 | 2 | .857 | —— | .857 | —— | 1 | 0 | 1 | 0 | 0 | .291 | —— | .291 | .857 | 3 | 0 | 3 | 1 | 5 |
| | t3 | 2 | 2 | 4 | 6 | —— | .291 | .744 | .998 | 0 | 3 | 3 | 1 | 1 | —— | .038* | .904 | .998 | 0 | 5 | 2 | 1 | 6 |
| | t4 | 2 | 2 | 4 | 4 | —— | .857 | .982 | .982 | 0 | 1 | 1 | 1 | 2 | —— | .570 | .982 | .982 | 0 | 2 | 1 | 1 | 9 |
| | t5 | 2 | 1 | 4 | 6 | .857 | .117 | .904 | —— | 1 | 4 | 2 | 0 | 2 | —— | .570 | .982 | .939 | 0 | 2 | 1 | 3 | 6 |
| | t6 | 2 | 1 | 3 | 6 | —— | .117 | —— | —— | 0 | 4 | 0 | 0 | 2 | —— | .038* | —— | .998 | 0 | 5 | 0 | 1 | 6 |
| | t7 | 2 | 2 | 4 | 6 | —— | .291 | —— | —— | 0 | 3 | 0 | 0 | 2 | .570 | .002** | .744 | .939 | 2 | 7 | 3 | 3 | 3 |
| | t8 | 2 | 2 | 4 | 4 | .117 | .570 | .313 | —— | 4 | 2 | 4 | 0 | 1 | .570 | .117 | .313 | .744 | 2 | 4 | 4 | 3 | 7 |
| | t9 | 2 | 1 | 4 | 2 | —— | —— | —— | —— | 0 | 0 | 0 | 0 | 1 | —— | —— | —— | —— | 0 | 0 | 0 | 0 | 2 |
| | t10 | 2 | 1 | 3 | 2 | .291 | —— | .744 | —— | 3 | 0 | 3 | 0 | 5 | .117 | .616 | .169 | —— | 4 | 1 | 6 | 0 | 12 |
| | t11 | 2 | 1 | 3 | 3 | .570 | —— | .904 | —— | 2 | 0 | 2 | 0 | 2 | .857 | —— | .904 | .787 | 1 | 0 | 2 | 2 | 7 |
| | | 24 | 19 | 45 | 46 | | | | | 15 | 17 | 20 | 2 | 20 | | | | | 40 | 42 | 56 | 17 | 69 |
| NONCOLLINEAR | t0 | 2 | 1 | 2 | 4 | .857 | .616 | —— | —— | 1 | 1 | 0 | 0 | 0 | .857 | .243 | —— | .982 | 1 | 2 | 0 | 1 | 5 |
| | t1 | 2 | 1 | 2 | 3 | .038* | .616 | —— | .948 | 5 | 1 | 0 | 1 | 0 | .857 | .616 | —— | —— | 1 | 1 | 0 | 0 | 5 |
| | t2 | 2 | 1 | 2 | 2 | —— | —— | —— | —— | 0 | 0 | 0 | 0 | 0 | .570 | —— | .243 | —— | 2 | 0 | 2 | 0 | 2 |
| | t3 | 2 | 2 | 0 | 8 | .117 | .616 | —— | .989 | 4 | 1 | 0 | 3 | 0 | .291 | .616 | —— | .492 | 3 | 1 | 0 | 8 | 3 |
| | t4 | 2 | 1 | 2 | 7 | .570 | .616 | —— | .994 | 2 | 1 | 0 | 2 | 0 | .291 | .616 | —— | .919 | 3 | 1 | 0 | 4 | 2 |
| | t5 | 2 | 1 | 2 | 7 | .857 | .857 | —— | —— | 1 | 1 | 0 | 1 | 0 | —— | .857 | .616 | .973 | 0 | 1 | 1 | 3 | 5 |
| | t6 | 2 | 1 | 2 | 6 | .857 | —— | —— | .984 | 1 | 0 | 0 | 2 | 0 | —— | .857 | —— | .193 | 0 | 1 | 0 | 8 | 0 |
| | t7 | 2 | 2 | 0 | 7 | .857 | —— | —— | .818 | 1 | 0 | 0 | 5 | 2 | —— | —— | —— | .500 | 0 | 0 | 0 | 7 | 2 |
| | t8 | 2 | 2 | 0 | 6 | .010* | .857 | .038* | —— | 6 | 1 | 5 | 0 | 1 | .002** | —— | .038* | .984 | 7 | 0 | 5 | 2 | 2 |
| | t9 | 2 | 1 | 3 | 3 | —— | —— | —— | —— | 0 | 0 | 0 | 0 | 0 | .570 | —— | .570 | .948 | 2 | 0 | 2 | 1 | 0 |
| | t10 | 2 | 2 | 2 | 8 | .117 | —— | —— | .989 | 4 | 0 | 0 | 3 | 1 | .291 | —— | —— | .998 | 3 | 0 | 0 | 2 | 2 |
| | t11 | 2 | 1 | 1 | 3 | .570 | —— | .570 | .948 | 2 | 0 | 2 | 1 | 0 | .857 | —— | .570 | .787 | 1 | 0 | 2 | 2 | 1 |
| | | 24 | 16 | 18 | 64 | | | | | 27 | 6 | 7 | 18 | 4 | | | | | 36 | 29 | 19 | 38 | 29 |

| | | Day + 3 Binomial *p* | | | | Relocation Errors | | | | | Day + 7 Binomial *p* | | | | Relocation Errors | | | | |
|---|---|---|---|---|---|---|---|---|---|---|---|---|---|---|---|---|---|---|---|
| | | WS | CV | CH | OT | WS | CV | CH | OT | OUT | WS | CV | CH | OT | WS | CV | CH | OT | OUT |
| COLLINEAR | t0 | .010* | —— | .059 | .117 | 6 | 0 | 6 | 4 | 6 | .002** | —— | .001** | .117 | 7 | 0 | 9 | 4 | 3 |
| | t1 | .010* | —— | .029* | .313 | 6 | 0 | 8 | 4 | 3 | .000*** | —— | .001*** | —— | 8 | 0 | 11 | 0 | 7 |
| | t2 | .291 | —— | .291 | .570 | 3 | 0 | 3 | 2 | 3 | .117 | —— | .117 | .857 | 4 | 0 | 4 | 1 | 7 |
| | t3 | .570 | .038* | .324 | .841 | 2 | 5 | 5 | 4 | 9 | .857 | .038* | .744 | .690 | 1 | 5 | 3 | 5 | 11 |
| | t4 | —— | .291 | .744 | .904 | 0 | 3 | 3 | 2 | 7 | .570 | .570 | .530 | .982 | 2 | 2 | 4 | 1 | 6 |
| | t5 | .291 | .038* | .530 | .841 | 3 | 5 | 4 | 4 | 6 | .117 | .291 | .324 | .841 | 4 | 3 | 5 | 4 | 12 |
| | t6 | —— | .002** | —— | .984 | 0 | 7 | 0 | 2 | 6 | .857 | .000*** | .904 | .939 | 1 | 10 | 2 | 3 | 5 |
| | t7 | .857 | .000*** | .904 | .841 | 1 | 9 | 2 | 4 | 5 | .117 | .002** | .169 | .998 | 4 | 7 | 6 | 1 | 5 |
| | t8 | .117 | .002** | .148 | .530 | 4 | 7 | 5 | 4 | 8 | .038* | .000*** | .006** | .530 | 5 | 8 | 8 | 4 | 7 |
| | t9 | .857 | —— | .857 | —— | 1 | 0 | 1 | 0 | 3 | .117 | —— | .117 | —— | 4 | 0 | 4 | 0 | 4 |
| | t10 | .038* | —— | .324 | —— | 5 | 0 | 5 | 0 | 14 | .570 | —— | .744 | .857 | 2 | 0 | 3 | 1 | 13 |
| | t11 | .291 | .243 | .324 | .787 | 3 | 2 | 5 | 2 | 8 | .117 | .616 | .076 | .948 | 4 | 1 | 7 | 1 | 8 |
| | | | | | | 34 | 38 | 47 | 32 | 78 | | | | | 46 | 36 | 66 | 25 | 88 |
| NONCOLLINEAR | t0 | .570 | .243 | —— | .904 | 2 | 2 | 0 | 2 | 4 | .291 | .014* | —— | .530 | 3 | 4 | 0 | 4 | 5 |
| | t1 | .570 | .243 | —— | —— | 2 | 2 | 0 | 0 | 6 | .291 | .616 | —— | —— | 3 | 1 | 0 | 0 | 8 |
| | t2 | .857 | —— | .616 | .570 | 1 | 0 | 1 | 2 | 1 | .570 | .616 | .616 | .570 | 2 | 1 | 1 | 2 | 1 |
| | t3 | .291 | —— | .857 | .902 | 3 | 0 | 1 | 5 | 8 | .570 | .067 | —— | .492 | 2 | 3 | 0 | 8 | 8 |
| | t4 | .857 | .616 | .291 | .818 | 1 | 1 | 3 | 5 | 2 | .291 | .243 | .570 | .671 | 3 | 2 | 2 | 6 | 2 |
| | t5 | .857 | .291 | .616 | .973 | 1 | 3 | 1 | 3 | 6 | .570 | .117 | —— | .919 | 2 | 4 | 0 | 4 | 7 |
| | t6 | .857 | —— | —— | .193 | 1 | 0 | 0 | 8 | 0 | .570 | .570 | —— | .333 | 2 | 2 | 0 | 7 | 2 |
| | t7 | .857 | —— | .857 | .500 | 1 | 0 | 1 | 7 | 5 | —— | —— | .570 | .500 | 0 | 0 | 2 | 7 | 5 |
| | t8 | .291 | .291 | .117 | .984 | 3 | 3 | 4 | 2 | 5 | .570 | .570 | .570 | .193 | 2 | 2 | 2 | 8 | 8 |
| | t9 | .291 | —— | .291 | —— | 3 | 0 | 3 | 0 | 2 | .570 | —— | .291 | .948 | 2 | 0 | 3 | 1 | 2 |
| | t10 | .291 | .616 | —— | —— | 3 | 1 | 0 | 1 | 5 | .570 | —— | —— | .902 | 2 | 0 | 0 | 5 | 7 |
| | t11 | .291 | —— | .291 | .787 | 3 | 0 | 3 | 2 | 3 | .038* | .616 | .117 | .546 | 5 | 1 | 4 | 3 | 2 |
| | | | | | | 36 | 29 | 19 | 37 | 47 | | | | | 28 | 20 | 14 | 55 | 57 |

t## = target ID; WS = error within set of three; CV = vertical collinear relocation error; CH = horizontal collinear relocation error; OT = other within-circle relocation error; OUT = outside circle relocation error; *$p < .05$; **$p < .01$; ***$p < .001$

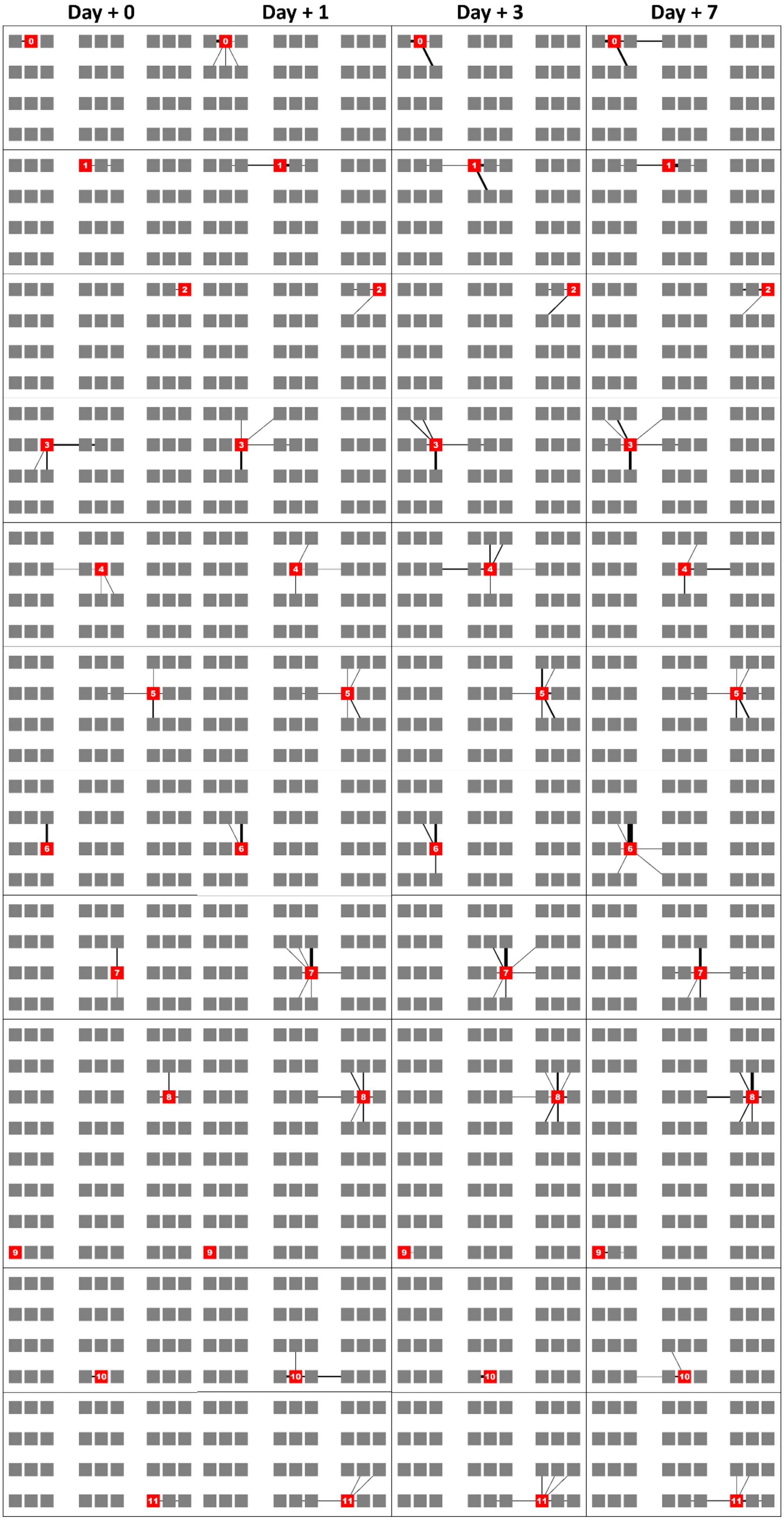

Table X. Experiment 3 within-circle errors progress by day, collinear condition.
Line weights depict number of errors to the distractor.

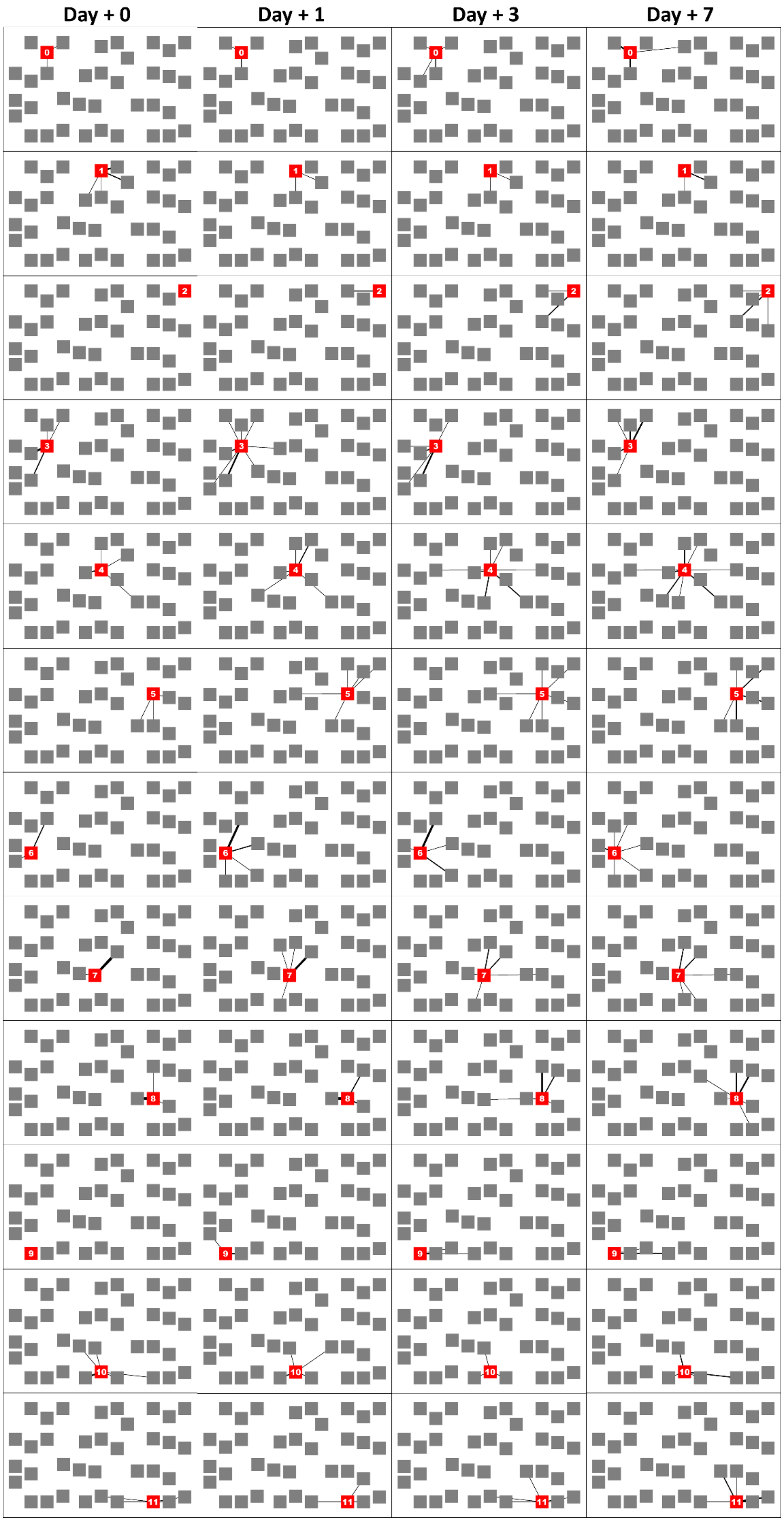

Table X. Experiment 3 within-circle errors progress by day, noncollinear condition.
Line weights depict number of errors to the distractor.

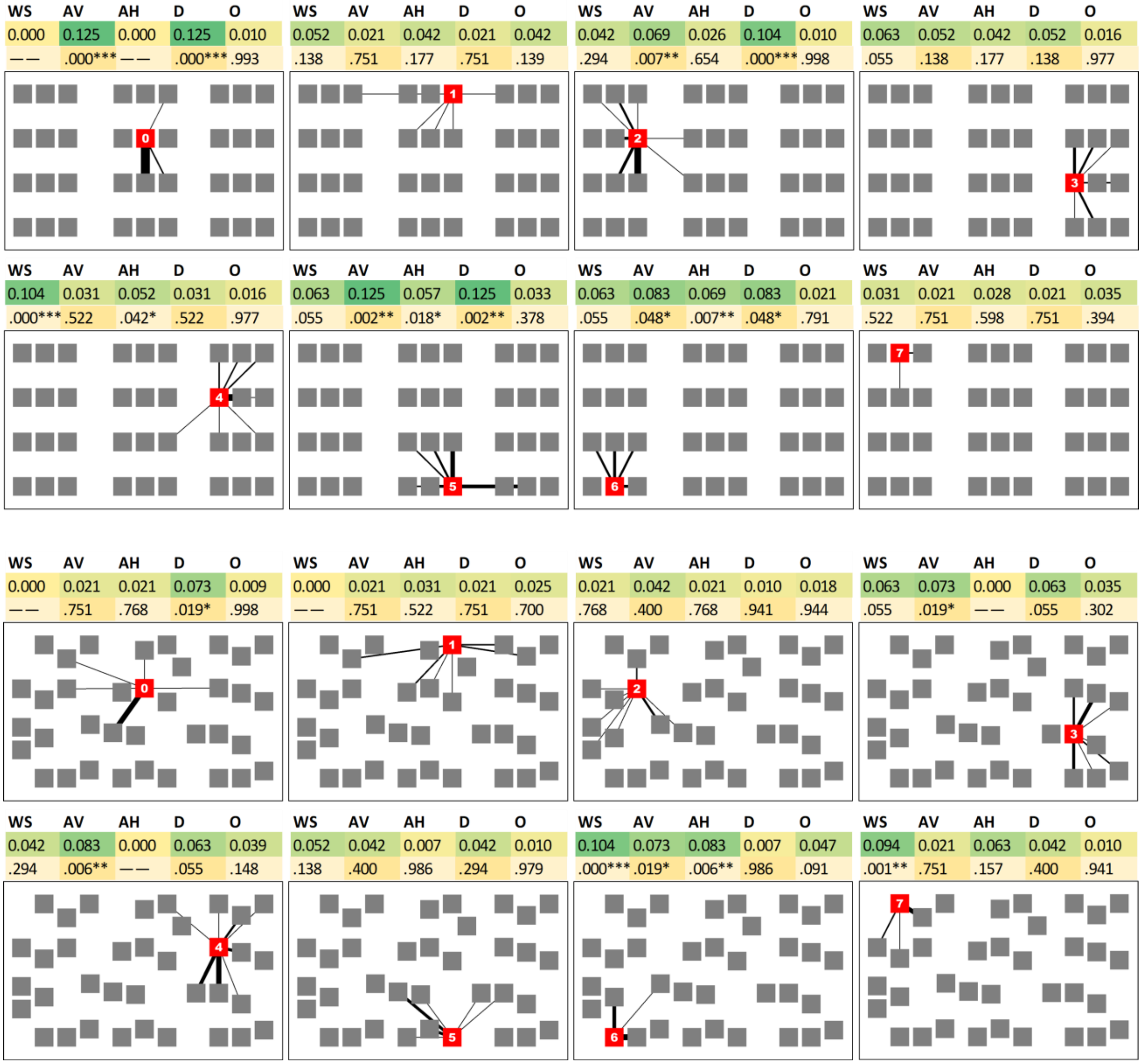

Figure 15. Experiment 4, Visualization of Specific Errors
WS = Within Set, AV = collinear vertical, AH = collinear horizontal, D = different set, O = other
Upper rows of numbers indicate binomial test results (yellow low, green high),
lower numbers are *p* values.

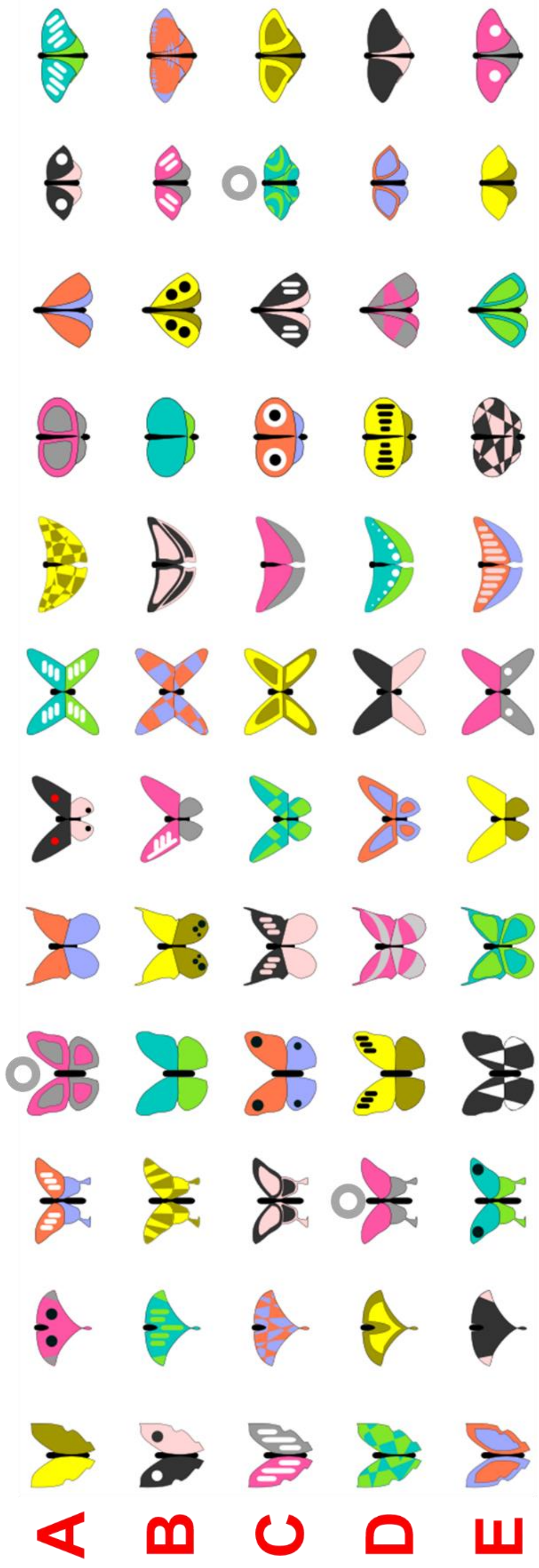

Figure 16. Five sets of systematically designed butterfly stimuli. Experiment 2 used grayscale, Experiment 3 used color. For Experiment 2, sets A, C, and D were used in balanced conditions to control for effects of design. Targets (small circles) were also carefully chosen to avoid excesses of size, coloration, and other feature saliency. For Experiment 3, sets A, B, and C were used as the 36 stimuli and these were also mixed and tested for effects of design, by rotating each set (A, B, or C) to serve as the target set of 12 and the other two sets to serve as the 24 distractors.